\documentclass[journal=jctcce,manuscript=letter,layout=traditional]{achemso} 
\usepackage[T1]{fontenc} 
\usepackage[version=3]{mhchem} 
\usepackage[dvipsnames]{xcolor}
\usepackage{graphicx,subcaption}
\usepackage{amssymb}
\usepackage{overpic}
\usepackage{braket}
\usepackage{comment}
\usepackage{soul}
\usepackage{bm}
\usepackage{hyperref}
\hypersetup{pdfborderstyle={/S/U/W 0.5}}
\usepackage{booktabs}
\usepackage{enumitem}

\newcommand{\hJ}{{\hat{J}}}

\newcommand{\bX}{{\bf X}}
\newcommand{\bP}{{\bf P}}
\newcommand{\bJ}{{\bf J}}

\newcommand{\bGamma}{{\bf \Gamma}}

\newcommand{\hbP}{{\hat{\bf P}}}
\newcommand{\hbJ}{{\hat{\bf J}}}
\newcommand{\hbd}{{\hat{\bf d}}}
\newcommand{\hbGamma}{{\hat{\bf \Gamma}}}

\newcommand{\bS}{{\bf S}}
\newcommand{\bL}{{\bf L}}

\newcommand{\REV}[1]{\textcolor{black}{#1}}
\author{Linqing Peng}
\email{lp9673@princeton.edu}
\affiliation{Department of Chemistry, Princeton University, Princeton, New Jersey 08544, United States}
\author{Tian Qiu}
\email{tianq@princeton.edu}
\affiliation{Department of Chemistry, Princeton University, Princeton, New Jersey 08544, United States}
\author{Nadine Bradbury}
\affiliation{Department of Chemistry, Princeton University, Princeton, New Jersey 08544, United States}
\author{Xuezhi Bian}
\affiliation{Department of Chemistry, Princeton University, Princeton, New Jersey 08544, United States}
\author{Mansi Bhati}
\affiliation{Department of Chemistry, Princeton University, Princeton, New Jersey 08544, United States}
\author{Robert Littlejohn}
\affiliation{Department of Physics, University of California, Berkeley, California 94720, United States}
\author{Nathanael M. Kidwell}
\affiliation{Department of Chemistry, The College of William and Mary, Williamsburg, Virginia 23187, United States}
\author{Joseph E. Subotnik}
\email{subotnik@princeton.edu}
\affiliation{Department of Chemistry, Princeton University, Princeton, New Jersey 08544, United States}
\title[]{Phase Space Electronic Structure Theory: From Diatomic Lambda-Doubling to Macroscopic Einstein-de Haas}

\begin{document}
\maketitle

\begin{abstract}
$\Lambda$-doubling of diatomic molecules is a subtle microscopic phenomenon that has long attracted the attention of experimental groups, insofar as rotation of molecular {\em nuclei} induces small energetic changes in the (degenerate) {\em electronic} state.
A direct description of such a phenomenon clearly requires going beyond the Born-Oppenheimer approximation. Here we show that a phase space theory previously developed to capture electronic momentum and model vibrational circular dichroism -- and which we have postulated should also describe the Einstein-de Haas effect, a macroscopic manifestation of angular momentum conservation -- is also able to recover the $\Lambda$-doubling energy splitting (or $\Lambda$-splitting) of the NO molecule nearly quantitatively and nonperturbatively (without a sum over states). The key observation is that, by parameterizing the electronic Hamiltonian in terms of both nuclear position ($\bX$) and nuclear momentum ($\bP$), a phase space method yields potential energy surfaces that explicitly include the electron-rotation coupling and correctly conserve angular momentum (which we show is essential to capture $\Lambda$-doubling). The data presented in this manuscript offer another small glimpse into the rich physics that one can learn from investigating phase space potential energy surfaces $E_{PS}(\bf X,\bf P)$ as a function of both nuclear position and momentum, all at a computational cost comparable to standard Born-Oppenheimer electronic structure calculations. 
\end{abstract}

\paragraph{Introduction}
Conservation of angular momentum is a fundamental principle that governs both everyday life and microscopic dynamics. As far as everyday life is concerned, angular momentum conservation leads to the Coriolis force that drives the clockwise subtropical Pacific Ocean gyre\cite{talley2011descriptive} in the Northern Hemisphere while also allowing a rolling hoop to stay upright\cite{kleppner2014introduction}. Because of angular momentum conservation,  whenever one part of a system changes its state of motion, the remaining degrees of freedom must respond so that the total momentum and angular momentum are preserved\cite{noether1971invariant}--a concept we are all well acquainted with.  On a microscopic level, the same logic applies to electrons and nuclei in molecules and solids: when the nuclei move, the electrons must adjust their states to the movement of the nuclei.
Just like Foucault pendulums, electrons can experience Coriolis forces.\cite{wick1948magnetic, van1951coupling,hehl1990inertial,serebrennikov2006coriolis,geilhufe2022dynamic}

Unfortunately, most modern-day electronic structure theories rely on the Born-Oppenheimer approximation and ignore the direct impact of nuclear momentum on electrons \cite{bian2023total}. To go beyond this limitation, we recently introduced a phase space theory that explicitly parametrizes the electronic Hamiltonian by both nuclear position and momentum and conserves the total momentum and angular momentum of electrons and nuclei.  We have already demonstrated that the method can recover vibrational circular dichroism\cite{duston2024phase,tao2024electronic} and Raman optical activity\cite{tao2025non}; for model systems, the approach can also offer improved vibrational energies\cite{bian2025phase,wu2025recovering}.  Moreover, because angular momentum conservation is guaranteed, we have postulated that the approach should be useful for modeling chiral-induced spin selectivity\cite{bian2025review} as well as the Einstein-de Haas (EdH) effect~\cite{einstein1915experimenteller, bradbury2025symmetry}.  The latter phenomenon 
remains of great interest to physicists and chemists alike\cite{mentink2019quantum, ganzhorn2016quantum}; as a brief review, EdH is the phenomenon whereby changing the magnetic moment can cause a ferromagnet to rotate, and thus serves as a macroscopically large example of angular momentum conservation in solids.

In the realm of small molecule chemistry, here we will now show that the same beyond BO physical concepts described above also naturally describe  $\Lambda$-doubling. In particular, using the phase space electronic structure theory from Refs.\citenum{qiu2024simple,tao2024practical,tao2025basis},  we will show that one naturally recovers the $\Lambda$-doubling splittings of the NO molecule quantitatively. Thus, we will argue that, over the smallest or largest length scales,  electronic structure theory can be improved by moving to a phase space framework where the Schrodinger equation is solved in the frame of moving nuclei.

An outline of this article is as follows.
First, for the sake of pedagogy, we will briefly review phase space electronic structure theory. Second, we will then discuss the physical meaning of phase space potential energy surfaces, showing that the energy surface minimum captures the strength of electron-rotation coupling for rotations around a given molecular axis. Third and finally, combined with a model of two-dimensional rotations, we will show how a phase space approach predicts the $\Lambda$-splitting of NO as observed in experiment.

As far as notation is concerned, as written below, boldface implies a 3- or 3N-dimensional vector. Operators are given hats. We will write the electronic angular momentum equivalently as $L_e$ or $L^e$ (depending on whether we have extra superscripts, e.g. $L_e^+$, or extra subscripts, e.g. $L^e_x$). Within phase space electronic structure theory, we label the canonical nuclear linear momentum $\bP$ and the angular momentum  $\bL$; however, as will be shown below, the canonical angular momentum within a semiclassical phase space theory represents the total angular momentum, so that for a system with electronic orbital and spin degrees of freedom, the total $\bL$ within a phase space calculation can be thought of as the total $\bJ$ within a fully quantum calculation.
The notation should be clear from context.

\paragraph{Brief Review of the Phase Space Method} \label{sec:review_PS}
Before we address $\Lambda$-doubling for NO, let us briefly review phase space electronic structure theory. Phase space electronic structure theory provides a semiclassical means to go beyond the standard BO approximation
by parametrizing the electronic Hamiltonian using both the classical nuclear position ($\mathbf{X}$) and nuclear momentum ($\mathbf{P}$).
More than ten years ago, Shenvi wrote down the first such example~\cite{shenvi2009phase}, suggesting that we diagonalize:
\begin{equation}
    \hat{H}_\mathrm{Shenvi}(\mathbf{X}, \mathbf{P}) = \sum_{A,IJK}\frac{1}{2M_A} (\mathbf{P}_A \delta_{IJ} - i\hbar \mathbf{d}^A_{IJ}) \cdot (\mathbf{P}_A \delta_{JK} - i\hbar \mathbf{d}^A_{JK}) | \Phi_I \rangle \langle\Phi_K| + \sum_I E_I |\Phi_I\rangle \langle\Phi_I| \label{Eq:shenvi}
\end{equation}
Here, $\bX$ and $\bP$ are classical c-numbers,  $\hat{H}_\mathrm{el}|\Phi_I\rangle = E_I|\Phi_I\rangle$ is the diagonalization of the standard electronic Hamiltonian, $\mathbf{d}_{IJ}^A = \langle\Phi_I | \frac{\partial}{\partial \mathbf{X}_A}|\Phi_J\rangle$, where $A$ is the atom index, is the derivative coupling vector between the many-body eigenstates $|\Phi_I\rangle$ and $|\Phi_J\rangle$, and $M_A$ is the mass of the atom $A$. 

Unfortunately, Eq.~\ref{Eq:shenvi} is of limited value for several reasons: (i) the derivative coupling can be numerically unstable near curve crossings, (ii) the derivative coupling is ambiguously defined for degenerate BO eigenstates, and  (iii) the procedure above requires two diagonalizations just for the electronic part and is computationally too demanding in practice. 
That being said, over the last several years, we have shown that progress can be achieved by approximating the derivative coupling vector $\mathbf{d}^A_{IJ}$ with the matrix element $\langle\Phi_I|\hat{\mathbf{\Gamma}}_A|\Phi_J\rangle$ of an one-body operator $\hat{\mathbf{\Gamma}}_A$.  To ensure that our approximation is as robust as possible, we have posited that one must preserve the analogs of four important symmetry constraints that derivative coupling vectors satisfy, namely:
\begin{gather}
    -i\hbar \sum_A \hat{\mathbf{\Gamma}}_A + \hat{\mathbf{p}} = 0, \label{sym1}\\
    \left[ -i\hbar{\sum_B \frac{\partial}{\partial \mathbf{X}_B}} + \mathbf{\hat{p}}, \hat{\mathbf{\Gamma}}_A\right] = 0, \label{sym2}\\
    -i\hbar \sum_A \mathbf{X}_A \times \hat{\mathbf{\Gamma}}_A + \hat{\mathbf{L}}^e + \hat{\mathbf{S}}^e = 0, \label{sym3}\\
    \left[ -i\hbar \sum_B \left(\mathbf{X}_B \times \frac{\partial}{\partial \mathbf{X}_B}\right)_\beta + \hat{l}_\beta + \hat{s}_\beta, \hat{\Gamma}_{A\gamma} \right] = i\hbar \sum_\alpha \epsilon_{\alpha \beta \gamma} \hat{\Gamma}_{A\alpha}. \label{sym4}
\end{gather} 
Eq.~\ref{sym1} and Eq.~\ref{sym3} reflect  translational invariance and rotational invariance of the total system, so that when nuclei are displaced translationally, the electrons  are displaced with them, and when nuclei rotate, the electrons rotate with them. Eq.~\ref{sym2} and Eq.~\ref{sym4} enforce that $\hat{\mathbf{\Gamma}}_A$ itself is invariant to translation and rotation. 

The form of the one-body operator $\hat{\mathbf{\Gamma}}_A$ we will use that satisfies all constraints above contains three terms, $\hat{\mathbf{\Gamma}}_A = 
\hat{\mathbf{\Gamma}}'_A +
\hat{\mathbf{\Gamma}}''_A + 
\hat{\mathbf{\Gamma}}'''_A$. To write down these operators, let us define a partition of unity of space according to nuclei as:
\begin{eqnarray}  
\label{eq:theta}
    \hat{\Theta}_A(\mathbf{x}) = \frac{Q_A e^{- |\hat{\mathbf{x}} - \mathbf{X}_A|^2/\sigma^2}}{\sum_B Q_B e^{- |\hat{\mathbf{x}} - \mathbf{X}_B|^2/\sigma^2}}
\end{eqnarray}
where $\sigma$ is a locality parameter in the unit of length.
The three $\bGamma$ operators are then:
$(i)$ The electron translation factor $\hat{\mathbf{\Gamma}}'_A$
\begin{gather}
    \hat{\mathbf{\Gamma}}'_A = \frac{1}{2i\hbar}(\hat{\Theta}_A \hat{\mathbf{p}} +\hat{\mathbf{p}} \hat{\Theta}_A),
\end{gather}
$(ii)$ the electron orbital rotation factor (ERF) $ \hat{\mathbf{\Gamma}}''_A$
\begin{gather}
    \hat{\mathbf{\Gamma}}''_A = \sum_B \zeta_{AB}(\mathbf{X}_A - \mathbf{X}_B^0) \times (\mathbf{K}_B^{-1} \hat{\mathbf{J}}^{(l)}_B), \\
    \hat{\mathbf{J}}_B^{(l)} = \frac{1}{2i\hbar}\left( (\hat{\mathbf{x}} - \mathbf{X}_B) \times (\hat{\Theta}_B \hat{\mathbf{p}})  + (\hat{\mathbf{x}} - \mathbf{X}_B) \times (\hat{\mathbf{p}}\hat{\Theta}_B ) \right),
\end{gather}
and $(iii)$ the electron spin rotation factor $\hat{\mathbf{\Gamma}}'''_A$
\begin{gather}
    \hat{\mathbf{\Gamma}}'''_A = \sum_B \zeta_{AB}(\mathbf{X}_A - \mathbf{X}_B^0) \times (\mathbf{K}_B^{-1} \hat{\mathbf{J}}^{(s)}_B), \\
    \hat{\mathbf{J}}_B^{(s)} = \frac{1}{i\hbar} \hat{\mathbf{S}}^e \hat{\Theta}_B, 
\end{gather}
where $ \mathbf{X}_B^0 = \frac{\sum_A\zeta_{AB} \mathbf{X}_A}{\sum_A \zeta_{AB}}$ is a local average position, 
\begin{equation}
    \mathbf{K}_B = \sum_A \zeta_{AB} \left((\mathbf{X}_A^\top \mathbf{X}_A - \mathbf{X}_B^{0\top} \mathbf{X}_B^0) \mathcal{I}_3 - (\mathbf{X}_A \mathbf{X}_A^\top - \mathbf{X}_B^0 \mathbf{X}_B^{0\top}) \right)
\end{equation}
is a local version of moment of inertia partitioned to each atom, and $\mathcal{I}_3$ is the $3\times3$ identity matrix. 
For the case of a linear molecule (as below) where $\mathbf{K}_B$ is not invertible, we replace $\mathbf{K}_B^{-1} \hat{\mathbf{J}}_B$ with~\cite{qiu2024simple}:
\begin{equation}
    \left(\sum_A \zeta_{AB} (\mathbf{X}_A-\mathbf{X}_B^0)^\top(\mathbf{X}_A-\mathbf{X}_B^0)\right)^{-1}(\mathcal{I}_3 - \mathbf{n}_3 \mathbf{n}_3^\top) \hat{\mathbf{J}}_B
\end{equation} 
where $\mathbf{n}_3$ is the unit vector along the internuclear axis of the linear molecule.  Above, $\zeta_{AB}=M_A e^{-|\mathbf{X}_A - \mathbf{X}_B|^2/8\sigma^2}$ reflects the degree of electronic/spin communication between different atoms and must be a function that decays on a length scale commensurate with the decay of the electronic density matrix.

Within our phase space electronic Hamiltonian, the $\hat{\mathbf{\Gamma}}_A$ operator   above creates an approximate Hamiltonian mimicking 
the derivative coupling:
\begin{align} \label{Eq:Hps}
    \hat{H}_\mathrm{PS}(\mathbf{X}, \mathbf{P}) =& \sum_{A}\frac{1}{2M_A} (\mathbf{P}_A - i\hbar \hat{\mathbf{\Gamma}}_A (\mathbf{X})) \cdot (\mathbf{P}_A  - i\hbar \hat{\mathbf{\Gamma}}_A (\mathbf{X})) + \hat{H}_\mathrm{el}(\mathbf{X}) \notag \\
    =&\sum_A \left(\frac{\mathbf{P}_A^2}{2M_A} - \frac{i\hbar}{M_A}\mathbf{P}_A \cdot \hat{\mathbf{\Gamma}}_A(\mathbf{X}) -\frac{\hbar^2}{2M_A}\hat{\mathbf{\Gamma}}_A^2 (\mathbf{X})\right) + \hat{H}_\mathrm{el}(\mathbf{X})
\end{align}
As with standard electronic structure, Eq.~\ref{Eq:Hps} can be solved self-consistently using a standard electronic structure solver with almost the same cost as the BO Hamiltonian. But in sharp contrast to the BO Hamiltonian, we now have an explicit electron-nuclear momentum coupling term, $-\sum_A \frac{i\hbar}{M_A}\mathbf{P}_A \cdot \hat{\mathbf{\Gamma}}_A(\mathbf{X})$, that provides the potential to capture various beyond BO effects. 

For example, when the symmetry constraints above are satisfied, the coupling term exactly captures the non-inertial Coriolis effect on electrons in the rotating frame of rigid nuclei. To verify this statement, imagine that we include $\hat{\mathbf{\Gamma}}_A'$ and $\hat{\mathbf{\Gamma}}_A''$ in the $\hat{\mathbf{\Gamma}}_A$ operator
(and thus satisfy a spinless version of Eq.~\ref{sym3}, i.e., we calculate the spatial part of the electrons in the body frame while the electronic spins are in the space-fixed frame); the relevant $\hat{\mathbf{\Gamma}}_A$ operator satisfies  $    -i\hbar \sum_A \mathbf{X}_A \times \hat{\mathbf{\Gamma}}_A + \hat{\mathbf{L}}^e = 0$ and conserves the total angular momentum excluding spin\cite{qiu2024simple}. When the nuclei are rotating around the origin (we set the center of mass as the origin) with an angular velocity $\omega$, the coupling term produced by the $\hat{\mathbf{\Gamma}}$  becomes
\begin{align}
        &-\sum_A \frac{i\hbar}{M_A}\mathbf{P}_A \cdot (\hat{\mathbf{\Gamma}}_A'(\mathbf{X})  + \hat{\mathbf{\Gamma}}_A''(\mathbf{X})) \notag \\
        =&   -\sum_A i\hbar \left(\mathbf{\omega} \times \mathbf{X}_A\right) \cdot (\hat{\mathbf{\Gamma}}_A'(\mathbf{X})  + \hat{\mathbf{\Gamma}}_A''(\mathbf{X})) \notag \\
        =&- i\hbar \mathbf{\omega} \cdot \sum_A \mathbf{X}_A \times (\hat{\mathbf{\Gamma}}_A'(\mathbf{X})  + \hat{\mathbf{\Gamma}}_A''(\mathbf{X})) \notag \\
        =& -\mathbf{\omega} \cdot \hat{\mathbf{L}}^e
\end{align}
which is exactly the electronic Coriolis term in a rotating non-inertial frame of an angular velocity $\mathbf{\omega}$. Previously, this result has been shown to hold generally for any type of nuclear motion, including both rotations and vibrations, in the nonlocal limit~\cite{tao2025basis,qiu2024simple} (where $\sigma \rightarrow \infty$ in Eq.~\ref{eq:theta}); here, we emphasize that this result holds regardless of the choice of the locality parameter in the case of pure rotations. Moreover, further including the electron spin rotation factor $\hat{\mathbf{\Gamma}}_A'''$ so as to satisfy the original Eq.~\ref{sym3} (corresponding to a calculation of the spatial {\em and} spin components of electrons in the body frame), we find a coupling term 
\begin{align}
       -\sum_A \frac{i\hbar}{M_A}\mathbf{P}_A \cdot \hat{\mathbf{\Gamma}}_A'''(\mathbf{X}) =- i\hbar \mathbf{\omega} \cdot \sum_A \mathbf{X}_A \times \hat{\mathbf{\Gamma}}_A'''(\mathbf{X})  = -\mathbf{\omega} \cdot \hat{\mathbf{S}}^e
\end{align}
which is a spin Coriolis term.  

Throughout this article, we will include all $\hat{\mathbf{\Gamma}}', \hat{\mathbf{\Gamma}}''$ and $\hat{\mathbf{\Gamma}}'''$ in $\hat{\mathbf{\Gamma}}$ unless otherwise noted. As we will show below, the features above enable us to describe the $\Lambda-$doubling effect -- an effect traditionally considered inaccessible directly within a BO approximation.

\paragraph{Phase Space in the Case of Degeneracy}

One of the most interesting facets of phase space electronic structure theory is the behavior of potential energy surfaces in the presence of degeneracy. In this context, a phase space approach to electronic structure can break symmetries that are protected within BO theory. In particular, with the BO framework, every radical molecule (with one unpaired electron)  must be part of a degenerate electronic subspace (usually a doublet) according to  Kramers' theorem\cite{kramers1930theorie}. That being said, however, according to a phase space description, the two electronic states may interact differently with nuclear motion (that breaks time reversal symmetry) and lose their energetic degeneracy.  As a result, even for the smallest molecule, one finds multiple minima in $\bP$-space; see Ref.~\citenum{bradbury2025symmetry}.

\subparagraph{Multiple Minima on the Potential Energy Surfaces} \label{sec:L_is_J}

\begin{figure}
    \centering
    \includegraphics[width=0.5\linewidth]{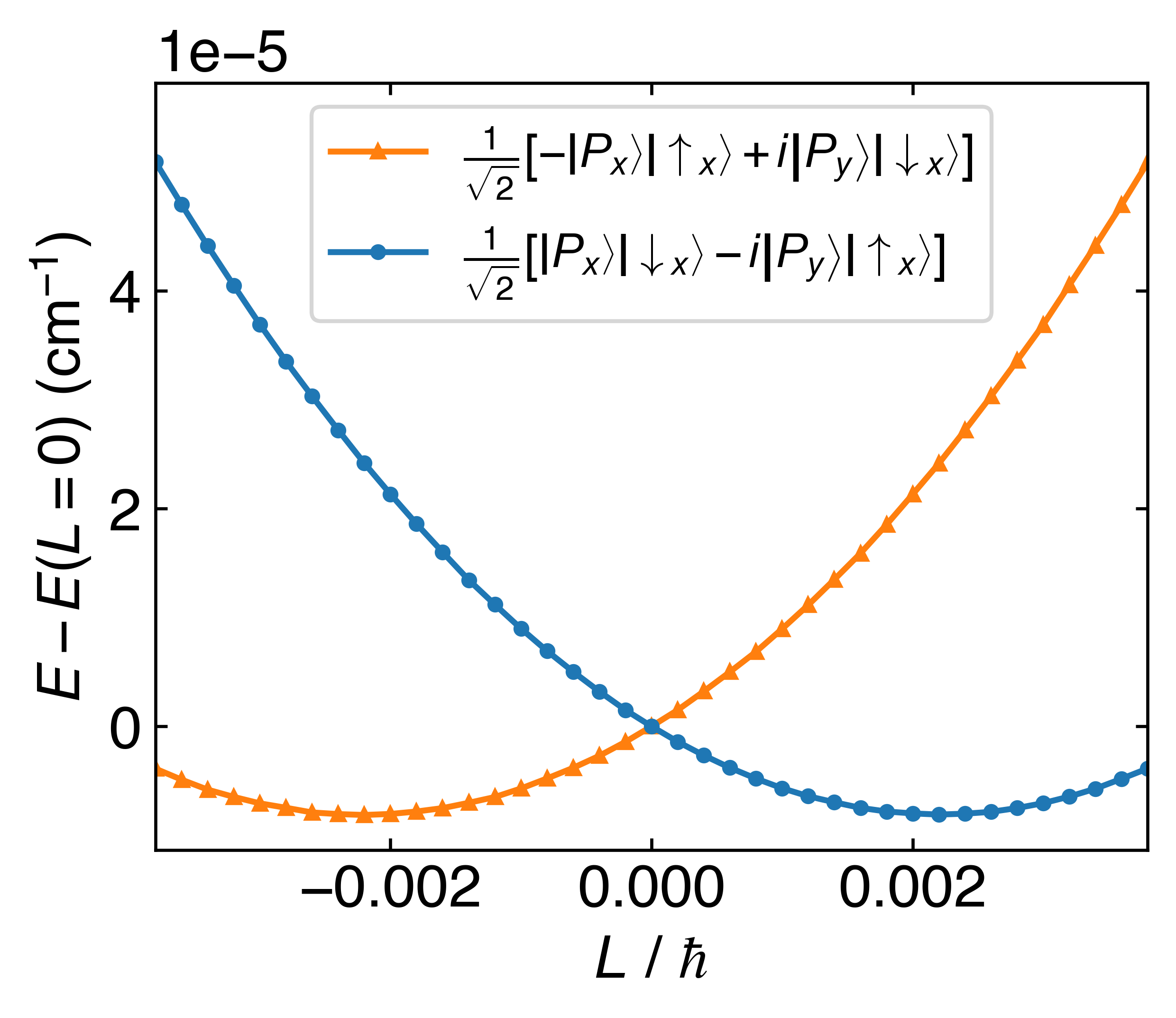}
    \caption{The potential energy surfaces of the ground and first excited states of NO from the phase space Hamiltonian, plotted versus the canonical nuclear angular momentum $L$. The blue and orange curves correspond to states whose valence $\Pi^*$ molecular orbital are approximately $\frac{1}{\sqrt{2}}\left[\left| P_x\right\rangle \left| \downarrow_x\right\rangle - i\left| P_y\right\rangle \left| \uparrow_x\right\rangle \right]$ and $\frac{1}{\sqrt{2}}\left[-\left|P_x\right\rangle \left| \uparrow_x\right\rangle + i\left|P_y\right\rangle \left| \downarrow_x\right\rangle \right]$, respectively. The nuclear coupling breaks the Kramers' degeneracy of the two PESs at a given finite canonical nuclear rotation angular momentum $L$ and form a double well with minima at $L/\hbar \approx \pm 0.0022$. } 
    \label{fig:well} 
\end{figure}

In Fig.~\ref{fig:well}, we plot the two lowest phase space surfaces from solving Eq.~\ref{Eq:Hps} for the molecule NO in a STO-3G basis. 
On the $x$-axis, our order parameter is the canonical nuclear rotation angular momentum $\mathbf{L} = \sum_A \mathbf{X}_A \times \mathbf{P}_A$.
Note that, whereas we find 
a doubly degenerate ground state PES under the BO framework, PS theory now splits into two shifted, approximately parabolic phase space PESs.  As will be discussed below in Sec.~\ref{Sec:literature}, the new broken-symmetry eigenstates have different parities in the case of diatomics.  Note, however, that the two PESs are still degenerate at zero canonical nuclear momentum and have a mirror symmetry with respect to $\mathbf{L}=0$, all due to  time reversal symmetry in the combined electronic and nuclear space. 

In order to understand the physical meaning of the two split PESs and of their unique minima along the canonical nuclear rotational angular momentum $\mathbf{L}$, one must recognize that the variable $\mathbf{P}$ -- which would appear  to be the nuclear linear momentum within the Born-Oppenheimer framework -- represents instead the combined nuclear and electronic momentum~\cite{littlejohn2024diagonalizing}. This representation can be shown by applying the total momentum operator $\hat{\mathbf{P}} + \hat{\mathbf{p}} = \sum_A \hat{\mathbf{P}}_A + \sum_i\hat{\mathbf{p}}_i$ to a coupled electronic and nuclear wavefunction $\Psi(\mathbf{X}, \mathbf{\mathbf{r}}) = \sum_n \psi_n (\mathbf{X}) \phi_n(\mathbf{X}, \mathbf{\mathbf{r}})$ in the BO formalism 
\begin{align}
    (\hat{\mathbf{P}} + \hat{\mathbf{p}})\Psi(\mathbf{X}, \mathbf{\mathbf{r}}) = & \sum_n \left[(\hat{\mathbf{P}} + \hat{\mathbf{p}}) \psi_n (\mathbf{X})\right]\phi_n(\mathbf{X}, \mathbf{r})  + \psi_n (\mathbf{X}) \left[(\hat{\mathbf{P}} + \hat{\mathbf{p}})\phi_n(\mathbf{X}, \mathbf{r}) \right] \notag \\
    =&\sum_n \left[(\hat{\mathbf{P}} + \hat{\mathbf{p}}) \psi_n (\mathbf{X})\right]\phi_n(\mathbf{X}, \mathbf{r})  \notag \\
    =& \sum_n \left(\hat{\mathbf{P}} \psi_n (\mathbf{X})\right)\phi_n(\mathbf{X}, \mathbf{r})
\end{align}
Here, we have used the fact that the electronic wavefunction $\phi_n(\mathbf{X}, \mathbf{\mathbf{r}})$ depends only on the relative electron-nucleus positions, 
\begin{equation} 
    \hat{\mathbf{P}} \phi_n = -\hat{\mathbf{p}}\phi_n = i\hbar \sum_i \nabla_i\phi_n. \label{en_deriv}
\end{equation}
Thus, the nuclear momentum evaluated on the nuclear wavefunction $\sum_n \langle\psi_n| \hat{\mathbf{P}} |\psi_n\rangle$ in fact represents the total momentum of the electrons and nuclei $\langle\Psi|( \hat{\mathbf{P}} + \hat{\mathbf{p}})|\Psi\rangle$. In fact, when one rewrites the Schrodinger equation in the adiabatic basis and replaces 
\begin{eqnarray}
    \hbP_A \rightarrow \hbP_A - i\hbar \hbd_A
\end{eqnarray}
where $\hbd_A$ is the derivative coupling operator, one can intuitively think of $\hbd_A$  as incorporating the electronic momentum\cite{littlejohn2023representation,littlejohn2024diagonalizing}. Thus, while $\hbP_A$ represents the total electronic + nuclear momentum,   $\hbP_A - i\hbar  \hbd_A$ represents the {\em nuclear} momentum. 
Analogous statements apply to 
angular momentum as well as linear momentum: 
the canonical nuclear rotational angular momentum $\mathbf{L}$ in fact represents the total angular momentum $\mathbf{J} = \mathbf{R} + \mathbf{L}^e + \mathbf{S}^e$  where $\mathbf{R}$ is the nuclear rotational angular momentum, $\mathbf{L}^e$ and $\mathbf{S}^e$ are the electronic orbital and spin angular momentum~\cite{vanHorn2025electronic}.

The facts above form the basis for phase space electronic structure theory.  As described above, within PS theory, we regard $\hbP_A$ as a classical object $\bP_A$, but we replace $\hbd_A$ with $\hbGamma_A$; thus, at the end of the day, the semi-classical quantity
\begin{eqnarray}
\label{eq:def:PiA}
    {\bf \Pi}_A \equiv \bP_A - i\hbar\hbGamma_A
\end{eqnarray}
represents the {\em nuclear} momentum and 
is known as the {\em kinetic} linear momentum. 
At this point, the physical meaning of the multiple minima of $E_\mathrm{PS}$ becomes clear. Namely, energy minima can arise only when the kinetic momentum of each nucleus is zero: 
\begin{equation}\label{eq:minPS}
    \frac{\partial E_\mathrm{PS}}{\partial \mathbf{P}_A} = \frac{\mathbf{P}_A - i\hbar\langle \hat{\mathbf{\Gamma}}_A\rangle}{M_A} = 0
\end{equation}
Thus, at the minimum, the value of $\mathbf{P}_A^\mathrm{min}$ is dictated by the different (nonzero) electronic momenta that can be captured by the different electronic states in a degenerate subspace,  $\mathbf{P}_A^\mathrm{min} = i\hbar\langle \hat{\mathbf{\Gamma}}_A\rangle$. Again, analogous statements hold for angular momentum, $\mathbf{L}^\mathrm{min}_A = i\hbar \mathbf{X}_A \times \langle \hat{\mathbf{\Gamma}}_A\rangle$, which is more rich because these angular momenta couple with each other and with the environment as magnetic dipoles.

\subparagraph{Extracting Rotational Energies}
\label{sec-rot_app}
Within BO theory, the simplest means of extracting vibrational and rotational energies is very well explored\cite{barone2011computational,wilson1980molecular,mccoy1992algebraic}.  Namely, one finds the local minimum geometry $\bX_0$ of $E_{BO}(\bX_0)$ in nuclear configuration space, one separates rotations from translations by looking for zeros of the Hessian, one makes a rigid rotor approximation for the rotations, and one makes a harmonic approximation for the vibrations.

How should one accomplish this task within a phase space perspective, given the existence of multiple minima in $E_{PS}(\bX,\bP)$? With reference to Eq. \ref{Eq:Hps} above, the  simplest means is to  diagonalize the electronic hamiltonian, and then both make a harmonic approximation for $\hat{H}_\mathrm{el}$ around and fix the value of $\hbGamma$ at the Born-Oppenheimer equilibrium geometry $\bX_0$ (with $\bP = 0$):
\begin{align} \label{eq:easy_ps}
    \hat{H}_\mathrm{PS}^\mathrm{harm}(\mathbf{X}, \mathbf{P}) =& \sum_{A}\frac{1}{2M_A} (\hbP_A - i\hbar \hat{\mathbf{\Gamma}}_A (\mathbf{X}_0)) \cdot (\hbP_A  - i\hbar \hat{\mathbf{\Gamma}}_A (\mathbf{X}_0)) + \hat{H}_\mathrm{el, diag}^\mathrm{harm}(\mathbf{X}- \bX_0) 
\end{align}
Eq.~\ref{eq:easy_ps} is obviously quite reminiscent of more advanced theories of molecular dynamics that separate rotations and vibrations; for example, see Ref.~\citenum{bunker2006molecular}.
For this article (with a focus on the spectrum of NO), we will focus exclusively on rotations; vibrations and vibrational-rotational coupling will be treated within a forthcoming article.
With the  interpretation above, the rotational kinetic energy of a molecular system is written as 
\begin{equation} \label{Eq:Erot}
    E_\mathrm{rot} = \frac{\langle\mathbf{R}^2\rangle}{2I} = \frac{\langle(\mathbf{J} - \mathbf{L}^e -  \mathbf{S}^e)^2\rangle}{2I} = \frac{\langle(\hbJ - \hbJ^e)^2\rangle}{2I},
\end{equation}
so that one might conjecture a natural nuclear-electronic coupling  of the form: 
\begin{equation} \label{Eq:Ecoup}
    E_\mathrm{coupling} = I^{-1} \langle \hbJ \cdot \hbJ^e \rangle.
\end{equation}
The operator evaluated in Eq.~\ref{Eq:Ecoup} is often termed the $L$-uncoupling and $S$-uncoupling operators\cite{lefebvre2004spectra}. In the context of ground state diatomics, as we will show below (Sec. \ref{sec:2D_model}), such an electron-nuclear coupling exactly gives rise to the $\Lambda$-doubling effect.

\paragraph{\texorpdfstring{$\Lambda$-doubling and NO}{Lambda-doubling and NO}}

Strictly speaking (i.e. without the Born-Oppenheimer approximation), for a diatomic molecule, as for any  molecule, the only good quantum numbers are  $J$, the total angular momentum quantum number, and its lab-frame $z$-axis projection $M$;
each energy eigenstate can be simultaneously labeled by one $J$ and one $M$. 
The $\Lambda$-doubling effect occurs when there are two degenerate electronic configurations with positive and negative angular momentum projections $\Lambda$ along the internuclear axis according to the BO electronic Hamiltonian. As will be seen below, this degeneracy is lost due to electron-rotation coupling,  leading to a splitting in the rotational spectrum.

\subparagraph{ \texorpdfstring{Standard View of $\Lambda$-doubling}{Standard View of Lambda-doubling}}

\label{Sec:literature}
When the nuclear rotational coupling and spin-orbit coupling (SOC) are small compared to  electrostatic interactions, these objects are usually treated as perturbations on top of the nonrelativistic BO Hamiltonian $H_0$, i.e. 
\begin{eqnarray}
\label{eq:Hexpand}
H = H_\mathrm{0} + H_\mathrm{SOC} + H_\mathrm{rot}
\end{eqnarray}
In the case of a relatively strong SOC and weak rotational coupling  (Hund's case (a)), one expands the wavefunction in terms of a specific set of the eigenstates of the nonrelativistic BO Hamiltonian, where each function is labeled by the additional approximately good quantum numbers: the electronic orbital angular momentum projection along the internuclear axis $\Lambda$, the electronic spin angular momentum $S$, the latter's projection along the internuclear axis $\Sigma$, and $\Omega=\Lambda + \Sigma$ which is also the total $J$ projection along the internuclear axis (because any nuclear rotation angular momentum $\mathbf{R}$ has no projection along the internuclear axis). Within the BO approximation, each basis function is then a product of the electronic wavefunction, solved in the molecular frame defined by the nuclei, and a rotational wavefunction that describes the orientations of the molecule, including both nuclei and electrons, within the lab frame, 
\begin{eqnarray}
    |\psi(n ^{2S+1}\Lambda_\Omega; \nu JM)\rangle = |n \Lambda S \Sigma\rangle |\nu\rangle |J\Omega M\rangle.
\end{eqnarray} \label{eq:hunds_a_basis}
where $J$ is the total electronic plus nuclear angular momentum and $M$ its projection along the lab-frame $z$-axis, and we define:
\begin{eqnarray}
    \langle \phi, \theta, 0|J\Omega M\rangle = \left[\frac{2J+1}{4\pi}\right]^{1/2} D^{J*}_{\Omega M}(\phi, \theta, 0) . 
    \label{eq:D}
\end{eqnarray}
where $D$ is the Wigner $D$-matrix~\cite{Varshalovich1988}. Note that $J$ and $M$ are the only good quantum numbers for an isolated system that has rotational invariance. Additionally included in Eq.~\ref{eq:hunds_a_basis} is an energy label $n$ ($n=X, A, B ...$) and a separable vibrational wavefunction $|\nu\rangle$, but these parts will be omitted in the following to simplify the discussion. 

In the case of the NO molecule, the lowest energy states are primarily made up of the  $^{2}\Pi_{1/2}$ ($\Lambda = \pm 1$, $S=1/2$, $\Sigma=\mp 1/2$, $\Omega =\pm 1/2$) manifold; these states are fairly well separated from the higher energy $^{2}\Pi_{3/2}$ ($\Lambda = \pm 1$, $S=1/2$, $\Sigma=\pm 1/2$, $\Omega =\pm 3/2$)  states due to the strong SOC.
As a heteronuclear diatomic molecule with $C_{\infty v}$ point group symmetry, NO has a ``parity symmetry,'' or more precisely, a reflection symmetry with respect to any plane that contains its internuclear axis. 
Any eigenstate should preserve such symmetry (or in the case of degeneracy, one can choose eigenstates that preserve such symmetry), so within the $^{2}\Pi_{1/2}$ space, the basis functions come in pairs to form parity eigenstates
\begin{align} \label{J_wavefunction}
    |^2\Pi_{1/2}, J,M,p^\pm \rangle =& \frac{1}{\sqrt{2}} [|\Lambda=1,S=1/2, \Sigma=-1/2\rangle |J, \Omega=1/2, M\rangle   \notag\\
    &\pm |\Lambda=-1,S=1/2, \Sigma=1/2\rangle |J, \Omega=-1/2, M\rangle  ]. 
\end{align}
Here, $p^\pm$ refers to the parity eigenstate with total parity $\pm(-1)^{J-S}$ for electronic $\Pi$ terms. For the half-integer J values and $S=\frac{1}{2}$ considered here, the $|p^+\rangle$ state with parity $(-1)^{J-\frac{1}{2}}$ is further labeled $e$, while the $|p^-\rangle$ state with parity $-(-1)^{J-\frac{1}{2}}$ is labeled $f$. $A'$ and $A''$ have also been used for labeling based on the reflection symmetry of the electronic spatial coordinates.~\cite{alexander1988nomenclature} 

Now, if we consider the effect of nuclear rotations as in Eq. \ref{eq:Hexpand}, the  $^{2}\Pi$ manifold interacts with higher-lying $^{2}\Sigma$ states. (Importantly,  the $\Sigma$ label here signifies $\Lambda=0$; in an unfortunate aspect of small molecule notation, $\Sigma$ has two different meanings and the $\Sigma$ here is different from the $\Sigma$ in Eq. \ref{J_wavefunction} which refers to the internuclear axis projection of the spin angular momentum.)  For NO, the interaction with  high-lying $^{2}\Sigma$ states  involves the off-diagonal $L$-uncoupling operator\cite{lefebvre2004spectra}, 
\begin{eqnarray}
B(\hat{J}^+\hat{L}_e^- + \hat{J}^-\hat{L}_e^+)
\end{eqnarray}
where $B=1/2I$ is the rotational constant in the equilibrium geometry (divided by the factor of $\hbar^2$), and the off-diagonal SOC operator, $ A/2(\hat{L}_e^+\hat{S}_e^- + \hat{L}_e^-\hat{S}_e^+)$ where $A$ is the spin-orbit constant (also divided by the factor of $\hbar^2$). Altogether, this approach generates a sum-over-states first-order corrected ground-state wavefunction
\begin{equation}
    |\psi'\rangle = |0 \rangle + \sum_{i \neq 0} \frac{|i\rangle \langle i |\frac{A}{2}(\hat{L}_e^+\hat{S}_e^- + \hat{L}_e^-\hat{S}_e^+) + B(\hat{J}^+\hat{L}_e^- + \hat{J}^-\hat{L}_e^+)|0\rangle}{E_0 - E_i}
\end{equation}
where $|0\rangle$ is a BO ground state $|^2\Pi_{1/2}, J,M,p^\pm \rangle$ and each $|i\rangle$ is an excited state $|^2\Sigma_\mathrm{1/2}, J, M, p^\pm\rangle$ with the same parity as $|0\rangle$. The final sum-over-states second-order energy correction  is
\begin{equation}
    E^{(2)} = \sum_{i\neq 0} \frac{|\langle 0|\frac{A}{2}(\hat{L}_e^+\hat{S}_e^- + \hat{L}_e^-\hat{S}_e^+) + B(\hat{J}^+\hat{L}_e^- + \hat{J}^-\hat{L}_e^+)|i\rangle|^2}{E_{0} - E_i} \label{Eq:E2_PT}.
\end{equation}
Evaluating Eq.~\ref{Eq:E2_PT} is difficult as one needs to include enough states for convergence (which is many).
Finally, note that the $S$-uncoupling operator in Eq.~\ref{Eq:Ecoup} also couples together the $^2\Pi_{1/2}$ and $^2\Pi_{3/2}$ manifolds for $J\geq 3/2$. A careful diagonalization within the combined space of $^2\Pi_{1/2}$ and $^2\Pi_{3/2}$ manifolds~\cite{zare1988angular,hinkley1972lambda}  yields the following result for the $\Lambda$-splitting of the $^2\Pi_{1/2}$ manifold
\begin{equation}
    \Delta E = (J+\frac{1}{2}) \left[ ( 1- \frac{Y}{X} + \frac{2}{X})(\frac{1}{2}p +q) + \frac{2}{X} (J+\frac{3}{2})(J-\frac{1}{2})q \right] \label{Eq:dE_PT}
\end{equation}
where $Y=A/B$, $X^2=Y(Y-4)+4(J+\frac{1}{2})^2$, and
\begin{align} \label{Eq:pt_parameter}
    p =& 2\sum_{i\neq 0} \frac{\langle0|A \hat{L}_e^{+}|i\rangle \langle i | B \hat{L}_e^+|0\rangle}{E_0 - E_i} \\ \label{eq:p_parameter}
    q =& 2\sum_{i\neq 0} \frac{|\langle0|B \hat{L}_e^+|i\rangle|^2}{E_0 - E_i}
\end{align}
which are calculated by explicitly summing over all important $^2\Sigma$ excited states. 

This concludes our brief review of the standard, well-established formal theory of $\Lambda-$doubling\cite{brown2003rotational,zare1988angular,bunker2006molecular}. Note that below we will compare phase space theory directly to experimental data (rather than to Eqs.~\ref{Eq:dE_PT}-\ref{Eq:pt_parameter}), but the analysis above will be helpful for reasons of interpretation.
Note also that the derivation of Eq.~\ref{Eq:dE_PT} ignores the parity dependency of the energy denominator, which would require a much more involved analysis to include~\cite{gordon2025energy}.

\subparagraph{A 1D Model Hamiltonian Compatible with Phase Space Electronic Structure Theory} \label{sec:1D}

Eqs.~\ref{Eq:E2_PT}-\ref{Eq:pt_parameter}  above represent the standard, perturbative approach to $\Lambda$-doubling that one finds in the literature. We would now like to show that this result can also be recapitulated quite naturally (and without a sum over states) using PS theory; the PS approach is not based on perturbation theory. 

In line with Sec. \ref{sec-rot_app} above, the simplest model Hamiltonian that can capture coupled nuclear rotation-electronic effects is a
standard rigid rotor that can be rotated with angle $\theta$ around one axis (hereby the name ``1D'' model). We let $\theta$ be the rotational angle of interest and we imagine that there are two electronic states of interest (as in the case of NO described above in Fig.~\ref{fig:well}) with $\pm \alpha$ total electronic angular momentum. 
In the spirit of Eq. \ref{eq:def:PiA}, the kinetic momentum is $\frac{\hbar}{i}\frac{\partial}{\partial \theta} \pm \alpha$, which leads to the following  (trivially separable) Schrodinger equation:
\begin{eqnarray}
    \frac{1}{2I} \left(
    \begin{array}{cc}
    \left(\frac{\hbar}{i} \frac{\partial}{\partial \theta} - \alpha\right)^2 & 0 \\
    0 &     \left(\frac{\hbar}{i} \frac{\partial}{\partial \theta} + \alpha\right)^2
    \end{array}
    \right)  \psi = E \psi
\end{eqnarray}
Given the need for periodicity, the solutions to this Schrodinger equation are $\psi_n(\theta) = \exp(i n \theta)$ with energies
\begin{eqnarray}
    E_n = \frac{(n \hbar - \alpha)^2}{2I} \; \; \; \mbox{and} \; \; \;  E_n = \frac{(n \hbar + \alpha)^2}{2I}
\end{eqnarray}
where $n$ takes on half-integer numbers for systems with odd numbers of electrons and integer numbers for systems with even numbers of electrons.
The difference between these energetic solutions is:
\begin{eqnarray}
    \Delta E = \frac{2n\alpha\hbar}{I} \label{eq:1D_en}
\end{eqnarray}

This expression for 1D $\Lambda$-splitting can be further substantiated by comparing with the perturbative (non-phase space) result for a rotation $J$ in 1D. Without loss of generality, we look at a finite $J$ along the $a$-axis.  We label the eigenstates of the standard electronic Hamiltonian (without SOC) as $\ket{0}, \ket{1}, \ket{2} \ldots$; in cases of degeneracy between any two eigenstates, these states are rotated to form eigenstates with reflection symmetry. The $\Lambda$-splitting due to the second-order coupling between the SOC and $L$-uncoupling, which dominates \REV{in the $^2\Pi_{1/2}$ manifold }in molecules with strong SOC like NO, according to Eq.~\ref{Eq:E2_PT}, is
\begin{align}\label{eq:E_use_Le_ori}
    \Delta E =& \sum_{i\neq 0} \frac{4\langle 0|2B \hat{L}^e_a J_a | i\rangle \langle i| A (\hat{S}^e_a \hat{L}^e_a + \hat{S}^e_b \hat{L}^e_b)  |0\rangle}{E_0 - E_i} \notag \\
    = & 8BJ_a \langle0 |  \hat{L}^e_a | \Psi_\mathrm{SOC}^{(1)}\rangle  \notag \\
    \approx& 4BJ_a \langle\Psi_\mathrm{SOC}^{(1)}| \hat{L}^e_a | \Psi_\mathrm{SOC}^{(1)}\rangle 
\end{align}
where $| \Psi_\mathrm{SOC}^{(1)}\rangle$ is the first-order corrected wavefunction due to SOC when no nuclear rotation is considered
\begin{equation}
    |\Psi_\mathrm{SOC}^{(1)}\rangle = |0\rangle + \sum_{i\neq 0}\frac{ |i\rangle \langle i| A (\hat{S}^e_a \hat{L}^e_a + \hat{S}^e_b \hat{L}^e_b) | 0\rangle}{E_0 - E_i},
\end{equation}
Within a perturbative treatment, $| \Psi_\mathrm{SOC}^{(1)}\rangle$ should be a parity eigenstate. To a good approximation, we can also write Eq.~\ref{eq:E_use_Le_ori} above as 
\begin{eqnarray}\label{eq:E_use_Le}
    \Delta E & \approx & 4BJ_a\langle \hat{L}^e_a\rangle_{J_a=0} 
\end{eqnarray}
where $\langle \hat{L}^e_a\rangle_{J_a=0} 
\approx \langle\Psi_\mathrm{SOC}^{(1)} | \hat{L}^e_a | \Psi_\mathrm{SOC}^{(1)}\rangle $ is the expectation value of $\hat{L}^e_a$ with respect to a relativistic (SOC-including) BO eigenstate (again without any  nuclear rotation). Above we have used the fact that $\langle 0 |\hat{L}^e_a| 0\rangle = 0$.  

Eq.~\ref{eq:E_use_Le} matches Eq.~\ref{eq:1D_en} above if we set $\alpha = \langle \hat{L}^e_a\rangle_{J_a=0}$.

\subparagraph{A 2D Model Hamiltonian Compatible with Phase Space Electronic Structure Theory} \label{sec:2D_model}
Next, let us imagine using phase space electronic structure theory to describe $\Lambda$-doubling in a full three dimensional context, beyond the 1D model 
above (which is trivially solvable).  As we will see, a multidimensional model cannot be solved analytically and requires either a large matrix diagonalization or approximations.

To begin our multidimensionl analysis, let $c$ denote the molecular axis of the diatomic molecule (from N to O).
Due to the $C_{\infty v}$ symmetry of the NO molecule around the internuclear $c$-axis, any choice of an axis perpendicular to $c$ is equivalent. We will pick the $a$- and $b$-axes associated with $\chi=0$~\cite{zare1988angular,zare1973direct}, where $\chi$ is the last of the Euler angles $(\phi,\theta,\chi)$ that describes the molecular orientation in the $z$--$y$--$z$ convention.  
As can be found in many textbooks~\cite{bunker2006molecular,brown2003rotational}, the quantum mechanical Hamiltonian for the nuclear kinetic energy associated with rotations of a linear molecule
\begin{equation}\label{eq:diatomic_exact}
    \hat{H}_\mathrm{rot} = -\frac{\hbar^2}{2I}\left[ \frac{1}{\sin^2 \theta} \frac{\partial^2}{\partial\phi^2} + \frac{1}{\sin \theta} \frac{\partial}{\partial\theta} \left( \sin \theta \frac{\partial}{\partial\theta}\right) \right],
\end{equation}
where $I=I_a=I_b$, is isomorphic to the Hamiltonian 
\begin{equation}\label{eq:Hiso}
    \hat{H}_\mathrm{rot}^\mathrm{iso} = \frac{1}{2I} \left[ (\hat{J}_a - \hat{J}_a^e)^2 + (\hat{J}_b - \hat{J}_b^e)^2 \right] 
\end{equation}
The  Hamiltonian $\hat{H}_\mathrm{rot}^\mathrm{iso}$ has one more degree of freedom than the true Hamiltonian in Eq.~\ref{eq:diatomic_exact}; namely $\chi$ remains as an independent variable in the definition of $\hat{\mathbf{J}}$.
  For an isomophormism to hold, we must diagonalize Eq.~\ref{eq:Hiso} within a basis that imposes the condition  that $J_c = J_c^e$,
  so as to remove all extraneous solutions; for more details, see  Refs.~\citenum{brown2003rotational,zare1988angular}.

In the context of phase space electronic structure theory, we will use a variant of the above isomorphic Hamiltonian where the nuclei are classical. It is then clear that the ground state phase space potential energy surfaces for a diatomic can be constructed by minimizing the following Hamiltonian:
\begin{eqnarray}
    \hat{H}_\mathrm{rot} = \frac{(J_{a} - \hJ^e_a)^2 + (J_{b} - \hJ^e_b)^2}{2I} 
\end{eqnarray}
or in short hand (remembering that $I_c = 0$):
\begin{align}
    H_\mathrm{rot} =& \frac{(\mathbf{J} - \hat{\mathbf{J}}^\mathrm{e})^2}{2I} \notag \\
    =& \frac{(\mathbf{J})^2}{2I} - \frac{\mathbf{J} \cdot \hat{\mathbf{J}}^e}{I} + \frac{(\hat{\mathbf{J}}^e)^2}{2I}. \label{eq:H_rot}
\end{align}
Now, the first and third terms are merely constant energy shifts for the lowest two states, and thus we focus on only the coupling term.  Given that we are working with a system with two electronic states, we will use the form of the Pauli matrices $(\sigma_a,\sigma_b,\sigma_c$) for a spin 1/2 system and write: 
\begin{eqnarray} 
    \hJ^e_{\beta} = \alpha \hat{\sigma}_{\beta}, \mathrm{for}~ \beta = a,b,c 
\end{eqnarray}
where 

\begin{eqnarray}
\label{eq:alpha:norm}
    \alpha = \frac{1}{\sqrt{2}}||\hJ^e_a|| = \frac{1}{\sqrt{2}}||\hJ^e_b||.
\end{eqnarray}
and $||\hat{J}||$ denotes the 2-norm (square root of the sum of squares) of operator $\hat{J}$.
At this point, we can rewrite the coupling term as a matrix \REV{in the bases of $|\Lambda=1,S=1/2, \Sigma=-1/2\rangle$ and $|\Lambda=-1,S=1/2, \Sigma=1/2\rangle$ electronic states. Because we find the matrices of $\hat{J}^e_a$ and of $\hat{J}^e_b$ are real off-diagonal and complex off-diagonal in such bases with a certain phase choice, the coupling term becomes}
\begin{align}
    - \frac{\mathbf{J} \cdot \hat{\mathbf{J}}^e}{I} \rightarrow& - \frac{ \alpha  \left(J_a \hat{\sigma}_a + J_b \hat{\sigma}_b \right)}{I} \notag \\
    &= - \frac{\alpha}{I} 
        \begin{pmatrix}
        0 & J_a-iJ_b \\
        J_a+iJ_b & 0
        \end{pmatrix}
        \label{eq:ci:sortof}
\end{align}
and evaluate the energy splitting by diagonalizing the Hamiltonian
\begin{equation}
    \lambda =  \pm \frac{\alpha}{I} \sqrt{(J_a-iJ_b) (J_a+iJ_b)} = \pm\frac{\alpha}{I} \sqrt{\mathbf{J}^2}.
\end{equation}
In the last step, when evaluating $\mathbf{J}^2$, we have requantized  $\mathbf{J}$ so as to be the  total angular momentum operator $\hat{\mathbf{J}}$. The energy splitting becomes 
\begin{equation}  \label{eq:2D_en}
\Delta E =2|\lambda| = \frac{2\alpha}{I} \sqrt{J(J+1)}.
\end{equation}
Eq. \ref{eq:2D_en} is our most reliable result (fulfilling the expectation of Eq.~\ref{Eq:Ecoup}) and can be checked against experiment; in order to model $\Lambda$-doubling, in principle all we require are the matrices for $\hat{J}^e_a$ and $\hat{J}^e_b$ -- which can be calculated in several ways through an {\em ab initio} phase space framework.

\paragraph{{\em Ab Initio} Results} \label{sec:abinitio}

\begin{figure}[h]
    \centering
    \includegraphics[width=0.4\linewidth]{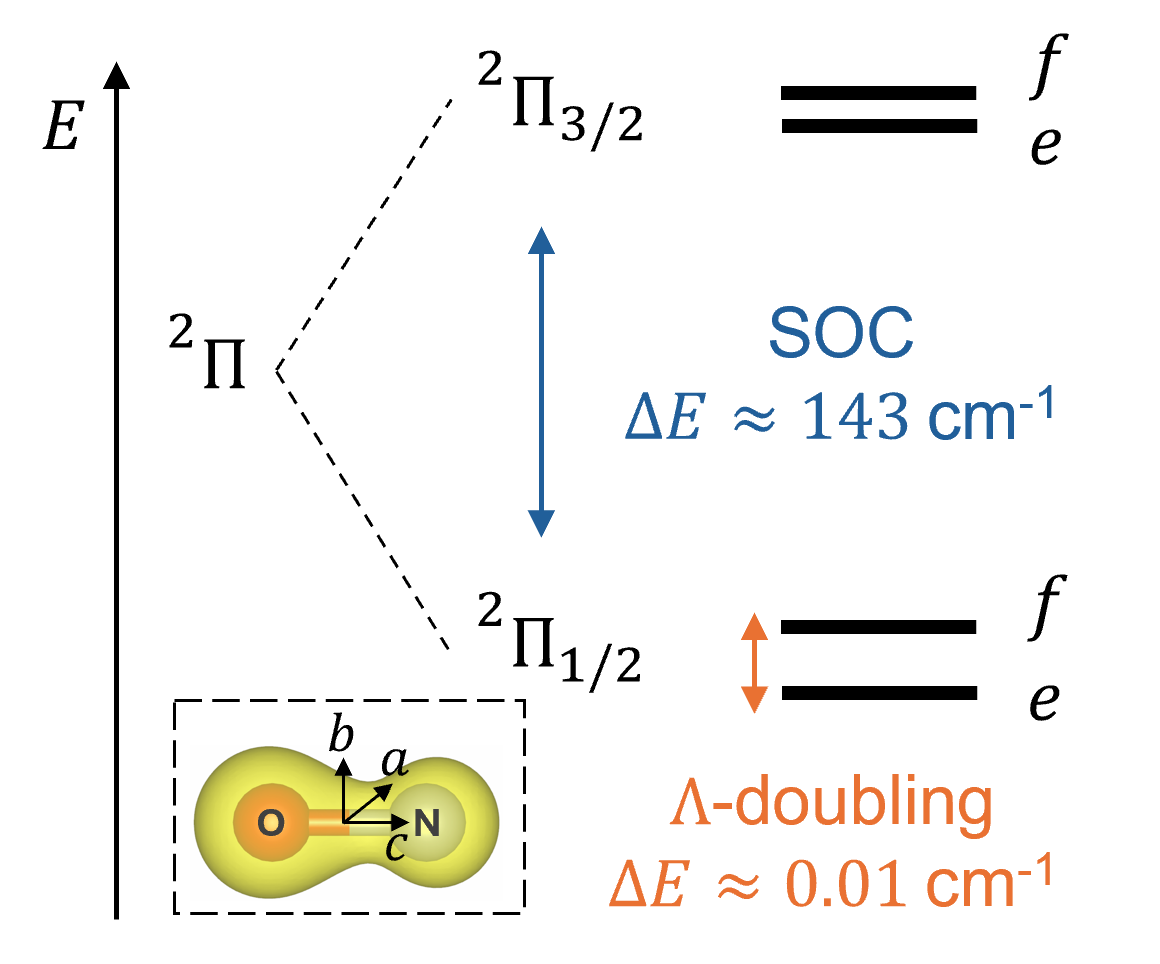}
    \caption{Energy diagram of the low-energy space of NO. Energy splittings are not drawn to scale. }
    \label{fig:E_diagram}
\end{figure}

Let us now address the necessary calculations. Below, we fixed the bond length of NO to be the experimental value~\cite{NIST_CCCBDB_10102439} 1.154 \AA ~and ignored dependence on vibrational and centrifugal distortion. Except where otherwise noted (see below), we treat SOC with the one-electron Breit-Pauli Hamiltonian~\cite{bethe2013quantum} (not including the two-electron spin-same-orbit and spin-other-orbit coupling).
In our study of NO,  as a sanity check, the first thing we calculated was the energy splitting between the $^2\Pi_{1/2}$ and $^2\Pi_{3/2}$ manifold at $L=0$ due to SOC.
For these calculations, we have run full configuration interaction (FCI) calculations for absolute certainty. 
 The low-energy eigenstates of NO, illustrated in Fig.~\ref{fig:E_diagram}, are split by strong SOC into the ground state $^{2}\Pi_{1/2}$ manifold and a $^{2}\Pi_{3/2}$ manifold that lies 143 cm$^{-1}$ higher, in reasonable agreement with the experimental separation~\cite{james1964spin} of 123 cm$^{-1}$ with about 16\% overestimation.   Note that this splitting is far bigger than the $\Lambda$-splitting; see Fig.~\ref{fig:E_diagram}.

Next, let us address  the $\Lambda$-splitting. \REV{In line with the discussion above, we will focus on the splitting of the ground doublet in the $^2\Pi_{1/2}$ manifold.}
As mentioned above,
if we wish to use the approach in Sec.~\ref{sec:2D_model}, there are two ways to extract 
the key objects, $\mathbf{L}^e$ and $\mathbf{S}^e$ (or really just $\mathbf{L}^e$ in practice), from a phase space calculation (and see Table~\ref{table:coupling} below.).

\begin{itemize}

\item Method I: On the one hand, one can  evaluate the two complete tensors in the low-energy manifold at $L=0$, i.e. one computes $\left(\hat{\mathbf{L}^e}+\hat{\mathbf{S}^e}\right)\big|_{L = 0}$, as suggested in Eq. \ref{eq:easy_ps} above and discussed also in Eq. \ref{eq:E_use_Le}.
Thereafter, one can set $\alpha = \frac{1}{\sqrt{2}} ||\hat{{L}^e_a}+\hat{{S}^e_a}||$.
Alternatively, for the NO problem, if one rotates to a basis with reflectional symmetry in the $b-c$ plane, $\hat{{L}^e_a}+\hat{{S}^e_a}$ will be diagonal and so we can set $\alpha = \left|\left<\hat{{L}^e_a}+\hat{{S}^e_a}\right>\big|_{L = 0} \right|$.

    \item Method II: 
On the other hand hand, if one  one seeks to intuitively extract rotational energies exclusively from the ground state phase space PES (without any further electronic calculations), one can alternatively find the PES minimum in phase space ($L_\gamma^\mathrm{min}$) as a function of rotation in each $\gamma$-direction. Because of Eqs. \ref{sym3} and \ref{eq:minPS} above, one is guaranteed that:
\begin{equation}
     L_\gamma^\mathrm{min} = \langle \hat{L}^e_{\gamma}+\hat{S}^e_{\gamma}\rangle_{L^{\mathrm{min}}_{\gamma}} 
\end{equation}
 Thus, given that $\hat{{L}^e_a}+\hat{{S}^e_a}$ will be diagonal in the ground state with the proper stymmetry (as noted for  Method I above), we can set $\alpha = L_\gamma^\mathrm{min}$.
Note that, for a nonlinear molecule with multiple different moments of inertia, this method would require several different optimizations to recover all three rotational directions.   Note also that, in practice,
the phase space minimum
$L_{\gamma}^\mathrm{min}$ is very small and close to zero for NO, so that 
$\langle \hat{L}^e_{\gamma}+\hat{S}^e_{\gamma}\rangle_{L^{\mathrm{min}}_{\gamma}} \approx \langle \hat{L}^e_{\gamma}+\hat{S}^e_{\gamma}\rangle_{L=0}$.
\end{itemize}

\noindent Finally, note that there is also a third and direct semiclassical approach.

\begin{itemize}
    \item Method III:
We can simply run a PS calculation with $L=1/2\hbar$ and obtain the $\Lambda$-splitting directly from the energy gap between two PESs (Fig.~\ref{fig:well}). 
\end{itemize}

\noindent 
  In practice, we find that all three of these approaches yield very similar results, though our results for Method II are perhaps the least accurate because of the need for a high resolution, accurate potential energy minimum. Note also that, for the NO molecule, the matrix element $\alpha$ is exclusively from the $\hat{\bL}^e$ operator; the spin component $\hat{\bS}^e$ is negligible. 

\subparagraph{Estimate of Coupling From PS calculations} \label{sec:3methods}

\begin{figure}
    \centering
    \includegraphics[width=0.5\linewidth]{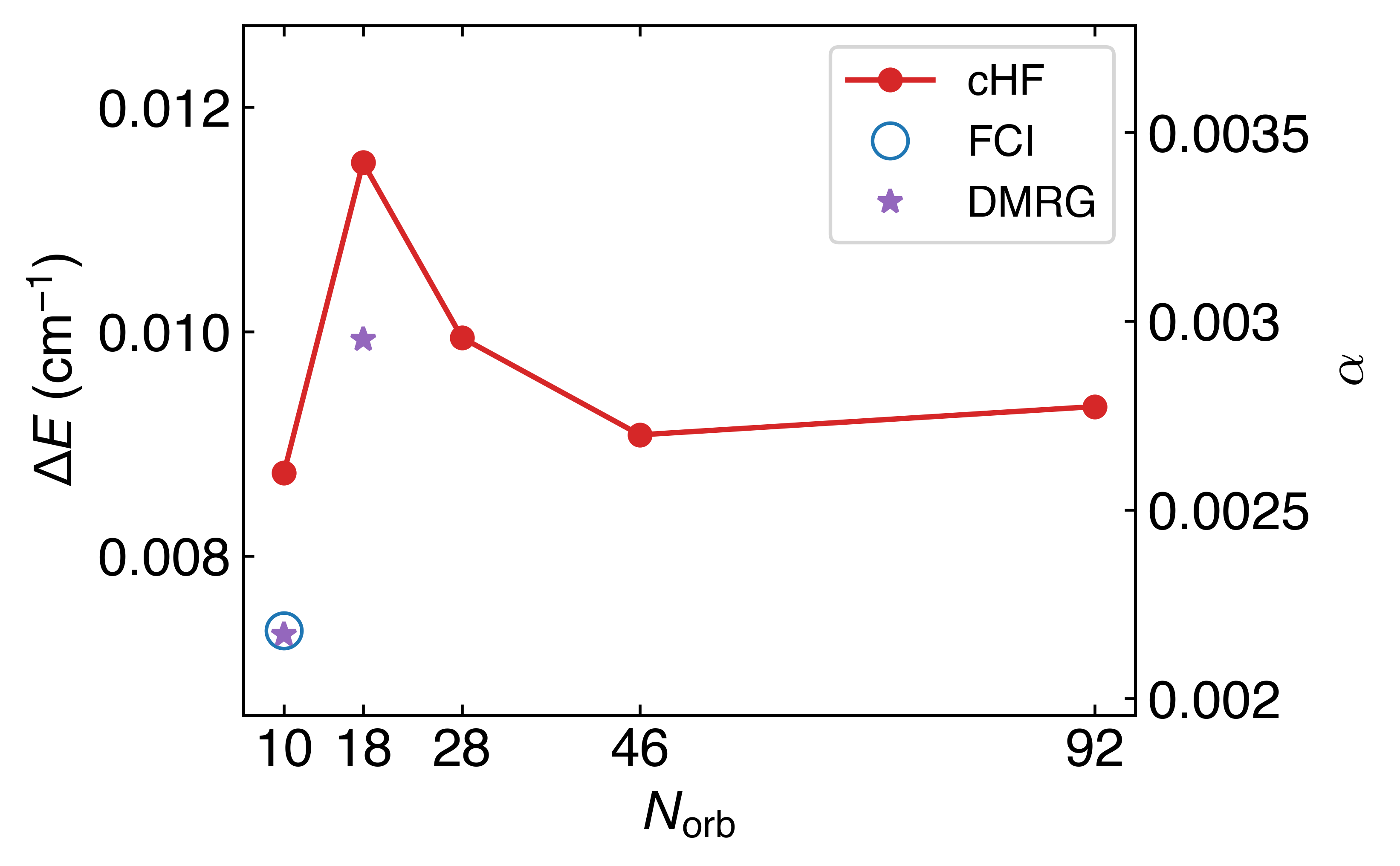}
    \caption{The $\Lambda$-splitting between the lowest $e$ and $f$ states and the associated $\alpha$ parameter for a nuclear rotation $J=1/2\hbar$ around one perpendicular axis. The results are calculated using the exact FCI solver (blue circle) in the STO-3G basis, the DMRG solver (purple star) which converges to the FCI accuracy in the STO-3G and 6-31G bases, and an approximate constrained Hartree-Fock solver in larger bases, from STO-3G, 6-31G, cc-pVDZ, aug-ccpVDZ, to aug-ccpVTZ (red).}
    \label{fig:1D_en}
\end{figure}

During the course of our investigation, our initial approach was to calculate the $\Lambda$-splitting directly by running a phase space electronic structure problem for a value of $L = \hbar/2$ (unless otherwise noted, and then directly measuring the gap). 
We performed several such  Method III calculations: 
\begin{enumerate}[label=(\roman*)]
    \item  We ran an FCI calculation in the minimal basis STO-3G.
    \item We solved the same Hamiltonian using the density matrix renormalization group (DMRG) solver~\cite{zhai2023block2} with enough bond dimension to converge to the FCI accuracy in a slightly larger basis 6-31G, after benchmarking against FCI in the STO-3G basis. For these calculations, the ground and first three excited states were solved using a state-averaged DMRG approach~\cite{dorando2007targeted}. Bond dimensions of 500 and 3000 were used for the STO-3G and 6-31G basis sets, respectively, to converge the first excitation energy to about $10^{-12}$ and $10^{-9}$ Ha precision. Unlike the calculations for (i) and (iii), for better numerical accuracy when evaluating a small energy gap, we invoked a linear approximation for the phase space energies and ran calculations with $L=3\hbar$ instead of $\hbar/2$ and rescaled the final energy splitting by 1/6; the assumption that the $\Lambda$-splitting scales approximately linearly with $J$ for small $J$ follows from the parabolic nature of Fig.~\ref{fig:well}.  
    \item In order to reach a full basis limit,  we developed and implemented a constrained Hartree-Fock (cHF)~\cite{ma2015constrained,kaduk2012constrained,peng2025accurate,behler2005dissociation} approach which avoids the incorrect broken symmetry solutions with $S^e_c=\pm 1/2$ and $L^e_c=\mp 1$ as found by  standard HF theory. Here, our constraint penalizes finite $\mathbf{S}^e$ vectors by adding an energy penalty term $\Delta E = -\lambda |\mathbf{S}^e|^2$ and therefore yields the correct parity eigenstates with $S^e_c, L^e_c\approx 0$. We applied this cHF method first with $\lambda = 0.1$, and then relaxed the converged solution with $\lambda = 0$ until convergence again. To obtain the first excited state with the opposite parity, we used the time reversal of the ground state electronic wavefunction as the initial guess, converged the cHF calculation with $\lambda = 0.1$, and then relaxed the solution with $\lambda = 0$. 
\end{enumerate}

All of our Method III results are shown in Fig.~\ref{fig:1D_en}. In the minimal basis, the first excitation energy from DMRG agrees with the FCI result within 0.5\% of error, confirming that the linear approximation of $\Lambda$-splitting holds quite  accurately and that the DMRG result in the 6-31G basis provides an accurate estimate of the FCI result. Relative  to the FCI and DMRG benchmarks, in both STO-3G and 6-31G basis sets, cHF consistently overestimates the energy splitting by about 20\%. Nevertheless,  cHF provides an efficient means to approach the complete basis set limit. The cHF results show that the basis-set incompleteness error in the minimal basis is small. Therefore, it is not unreasonable to expect that the minimal basis FCI result provides a good estimate of the FCI energy in the complete-basis limit. Thus, below we will use the FCI energy splitting in the STO-3G basis to estimate  
the $\alpha$ parameter in a 1D version of the model Hamiltonian Eq.~\ref{eq:H_rot} (``Method III'' in Table~\ref{table:coupling}).

As a complement to the Method III results, we also followed Method II above and 
scanned the FCI phase space PES looking for a minimum so as
to determine the $\alpha$ parameter based on Eq.~\ref{eq:alpha:norm}. We find that the PES minima are at $L_a=\pm0.0022$, and therefore, $\alpha = ||\hat{J}^e_a|| = 0.0022$ (``Method II'' in Table~\ref{table:coupling}). Lastly,
we also followed Method I above and computed the matrices of $\hat{\mathbf{L}}^e$ and $\hat{\mathbf{S}}^e$ for the two lowest energy eigenstates  at $\mathbf{L}=0$. This yields $\alpha=||\hat{J}^e_a|| = 0.0018$ (``Method I'' in Table~\ref{table:coupling}).
Thus, in total, we can estimate the splitting in three ways and they all yield consistent results.

\begin{table}   
  \centering
  \begin{tabular}{c c c c}
      \toprule
           Method & I & II & III \\ \midrule
           $\alpha$ & 0.0018 & 0.0022 & 0.0022  \\
              \bottomrule 
  \end{tabular}  
  \caption{The coupling parameter $\alpha$ estimated in three ways:  Method I: from $\langle \hat{L}^e_a + \hat{S}^e_a\rangle$ evaluated on the parity eigenstates at $\mathbf{L} = 0$; 
  Method II: from $L_a$ at the minimum of the PES; Method III: from energy splitting $\Delta E$ at $L_a = 1/2 \hbar$.  }
  \label{table:coupling}
\end{table}

Before concluding this section, we note that using $\hat{\mathbf{\Gamma}} = 
\hat{\mathbf{\Gamma}}' +
\hat{\mathbf{\Gamma}}''$ is enough; the absolute coupling is largely unchanged by adding $\hat{\mathbf{\Gamma}}'''$.  Both $\hat{\mathbf{\Gamma}}'$ and $\hat{\mathbf{\Gamma}}''$ contribute to the energy splitting.

\subparagraph{The Splitting of the $^2\Pi_{1/2}$ manifold as a function of \texorpdfstring{$J$}{J}}\label{sec:final_result}

Having shown that we can evaluate the $\Lambda$-splitting between the $e/f$ states for $J=1/2$ for a ``clamped'' rotation around one axis in several different consistently ways, we will now use the value of $\alpha$ found above and plug into Eq.~\ref{eq:2D_en} and make predictions for rotations in the full two dimensions and for larger $J$ values. In principle, one could use the 1D model in Eq.~\ref{eq:1D_en}, but we will use the 2D model in Eq.~\ref{eq:2D_en} (as one can always argue about the validity of a 1D rotational model).  That being said, in practice, the  only difference is a factor of $J$ versus $\sqrt{J(J+1)}$. 

\begin{figure}
    \centering
    \includegraphics[width=0.5\linewidth]{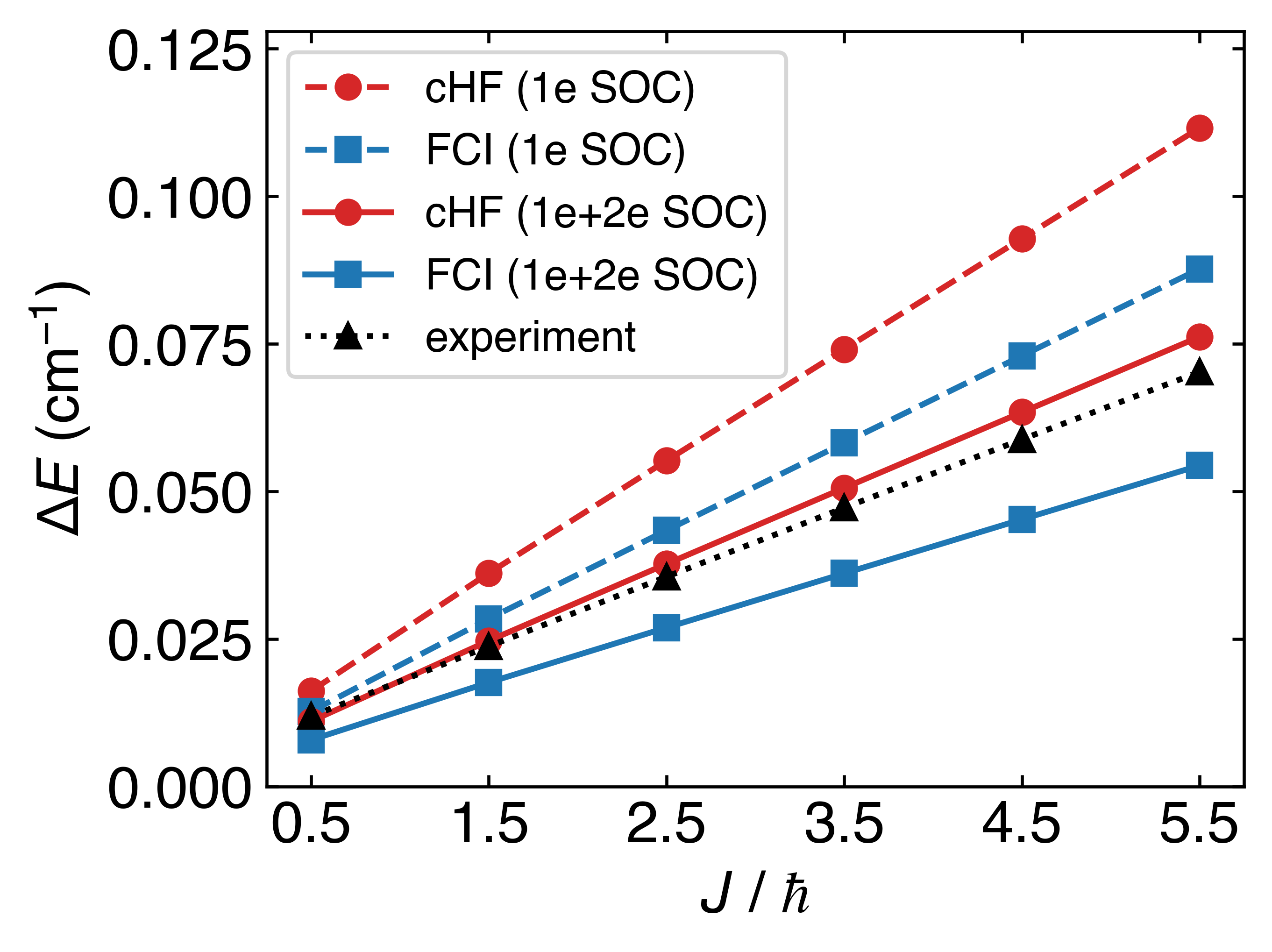} 
    \caption{Energy splitting between the $e$ and $f$ states as a function of total angular momentum $J$ from the 2D model parametrized with $\alpha$ derived from FCI in the STO-3G basis (blue) and from cHF in the aug-ccpVTZ basis (red), both using Method III, in comparison to experiments (black). Dashed lines include only the 1-electron SOC, while solid lines include both 1-electron and 2-electron SOC (with a spin-orbit mean-field approximation~\cite{neese2005efficient,lee2023ab}).}
    \label{fig:2D_en}
\end{figure}

In Fig.~\ref{fig:2D_en}, we compare the $\Lambda$-splitting hereby estimated with the experimental measurement for small half-integer $J$. Here, we plot FCI results using $\alpha$ estimated from the energy splitting in the minimal basis (Method III). When we only include the 1-electron SOC, these results are expected to moderately overestimate the splitting due to the overestimated SOC discussed at the beginning of Sec:~\ref{sec:abinitio}. We also plot cHF results in the complete basis set limit, which is also expected to overestimate the experimental answer, due to both the overestimation of SOC and the mean-field approximation.  According to Fig. \ref{fig:2D_en}, we find that if we include only the 1-electron SOC, the FCI results (``FCI (1e SOC)'' in Fig.~\ref{fig:2D_en}) overestimates the splitting by 7-25\% and the cHF results (``cHF (1e SOC)'') overestimate the splitting by 36-59\%. If we further include the 2-electron SOC under the spin-orbit mean-field approximation, the FCI results (``FCI (1e+2e SOC)'') underestimates the splitting by 34-23\% and the cHF results (``cHF (1e+2e SOC)'') deviate by only 4-9\%. It would appear that, if we include the full SOC tensor and work in a big enough basis, our semiclassical approach can be nearly quantitative--though further testing (and comparisons with fully quantum-mechanical calculations) will be necessary to say much more.

\paragraph{Discussion}

\subparagraph{Kramers' Theorem}

At this point, we have shown that a phase space interpretation of electronic structure theory automatically yields a meaningful $\Lambda$-splitting. Before we interpret this result, a few words are appropriate with regard to how our finding above relates to Kramers' theorem. 
Kramers' theorem states that for any time-reversal-invariant system with an odd number of electrons (such as NO), if we ignore nuclear spin, then every energy eigenstate is at least doubly degenerate, as each eigenstate $\ket{\psi}$ can be paired with its orthogonal time-reversal pair $T\ket{\psi}$. 
This statement holds when considering either (i) the electronic Hamiltonian within BO theory or (ii) the total electronic + nuclear Hamiltonian.  
For our purposes [focusing on (ii)], it is crucial to emphasize that the parity states (Eq.~\ref{J_wavefunction}) that are associated with $\Lambda$-splitting contain both electronic and rotational components. 
In what follows below, we will follow the phase convention of the time-reversal operator $T$ used by Condon and Shortley~\cite{condon1935theory} and by Hougen~\cite{hougen1970calculation}
\begin{gather}\label{eq:T_phase}
T(|J, \frac{1}{2}\rangle) = +|J, -\frac{1}{2}\rangle 
\end{gather}
where the second quantum number denotes either lab-$z$ or body-$c$ axis projection of $J$. The same convention is used for $|S, \Sigma\rangle$ and $|L, \Lambda\rangle$. Of course, in tandem with Eq.~\ref{eq:T_phase}, one must have:
\begin{gather}
T(|J, -\frac{1}{2}\rangle) = -|J, \frac{1}{2}\rangle 
\end{gather}

Let us now consider the implications  of Kramers' theorem for the wavefunctions of the low-energy manifold of NO. If we apply the time reversal operator to the electronic component, focusing on the dominant $^2\Pi_{1/2}$ contribution to the ground state, one finds:
\begin{eqnarray}
T(|\Lambda=\pm1,S=1/2, \Sigma=\mp1/2\rangle) = (-1)^{\Sigma-1/2+\Lambda}|\Lambda=\mp1,S=1/2, \Sigma=\pm1/2\rangle 
\end{eqnarray}
Next, if one applies time reversal to the rotational component, one finds:
\begin{eqnarray}
T(|J, \Omega=\pm1/2, M\rangle) = (-1)^{\Omega +M -1}|J, \Omega=\mp1/2, -M\rangle
\end{eqnarray} 
This result follows from Eqs. \ref{eq:T_phase} and \ref{eq:D} and the well known identities (for a half integer $M$):
\begin{eqnarray}
    D^J_{\Omega M}(\phi, \theta, \chi) &=& (-1)^{\Omega - M} D^J_{-\Omega, -M}(\phi, \theta, \chi)^* \\
    \Omega + M - 1 & = & \Omega - M + 2 (M- 1/2) \equiv  \Omega  - M
\mbox{\; (mod 2)}
\end{eqnarray}
As a result, if one applies the time reversal operator to the total (nuclear + electronic wavefunction in Eq. ~\ref{J_wavefunction}), one obtains:
\begin{align} 
    T(|^2\Pi_{1/2}, \nu, J,M,p^\pm \rangle) =& \frac{(-1)^{M-1/2}}{\sqrt{2}} [|\Lambda=-1,S=1/2, \Sigma=1/2\rangle |J, \Omega=-1/2, -M\rangle |\nu\rangle  \notag\\
    &\pm |\Lambda=1,S=1/2, \Sigma=-1/2\rangle |J, \Omega=1/2, -M\rangle  |\nu\rangle ] \notag \\
    =&\pm(-1)^{M-1/2}|^2\Pi_{1/2}, \nu, J,-M,p^\pm \rangle. \\
    T^2(|^2\Pi_{1/2}, \nu, J,M,p^\pm \rangle) =& -|^2\Pi_{1/2}, \nu, J,-M,p^\pm \rangle.
\end{align}
The same conclusion holds for the $^2\Sigma$ components in the low-energy state wavefunctions that the time reversal yields a state with the same parity but opposite $M$, apart from a phase factor. In the end, the time-reversed state is orthogonal to and degenerate with the original state; after all, the energy must be invariant to the sign of $M$ (which is really a statement about time-reversal symmetry). Of course, without an external field, there can be further symmetry. For example, without an external electric field, rotational invariance would imply not just two-fold degeneracy, but actually $2J+1$-fold degeneracy.  That being said, in the presence of an external electric field, two-fold degeneracy still remains (between $M$ and $-M$) for an odd spin system. In order to violate Kramers' theorem,  one must break time-reversal symmetry by, e.g., applying a magnetic field.

\subparagraph{A Phase Space Visualization of the e/f States}

Finally, how should we interpret the result above, where we have found a reasonably accurate estimate of $\Lambda$-doubling with a quite simple calculation?  
\begin{figure}
    \centering
    \includegraphics[width=1\linewidth]{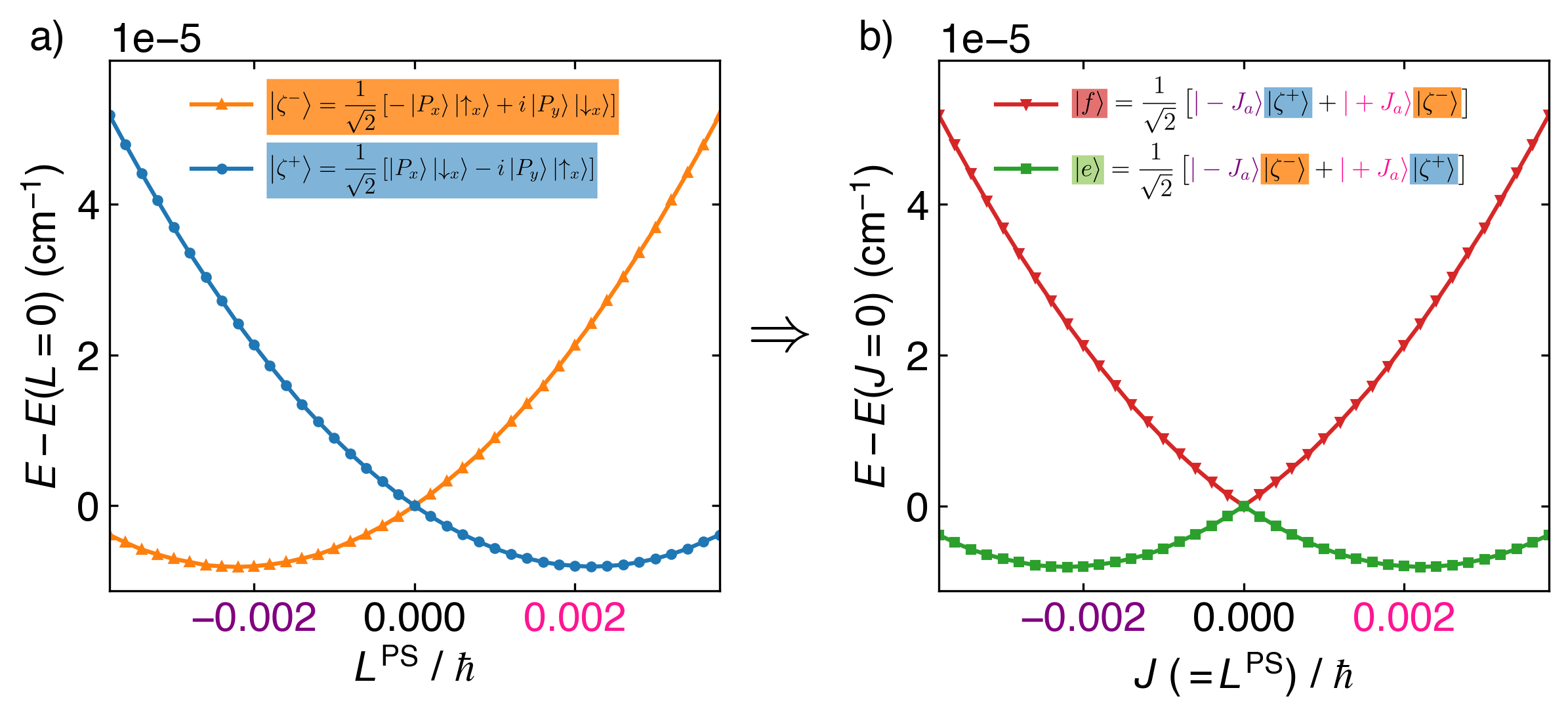}
    \caption{\REV{\textbf{a)} Our original view (see Fig.~\ref{fig:well}) of  potential energy surfaces of the ground and first excited states of NO according to a phase space electronic structure Hamiltonian as a function of the phase space canonical nuclear angular momentum $L$. \textbf{b)} The same data now labeled differently: The green and red curves can be understood as the $e$ and $f$ parity total electronic and nuclear states, now understood as functions of total angular momentum $J$ (represented by the phase space $L$). The $e/f$ states are the states that are observed spectroscopically. }}
    \label{fig:well_ef} 
\end{figure}
The phase space framework above paints a very intuitive picture for visualizing the $e$ and $f$ parity (total electronic and nuclear) wavefunctions in Eq.~\ref{J_wavefunction}. To establish such a picture, let us explore the phase space  PES for a rotation around  the $a$ axis, e.g.  $L_a> 0$. We  rewrite the dominant components of the low-energy eigenstate wavefunctions Eq.~\ref{J_wavefunction} as   
\begin{align}
    |^2\Pi_{1/2}, J,M,p^\pm \rangle = &\frac{1}{\sqrt{2}} [ \frac{1}{\sqrt{2}}(|J\Omega M\rangle + |J-\Omega M\rangle) \frac{1}{\sqrt{2}} (|\Lambda S - \Sigma \rangle \pm |-\Lambda S \Sigma \rangle)  \notag \\
    &+ \frac{1}{\sqrt{2}}(|J\Omega M\rangle - |J-\Omega M\rangle) \frac{1}{\sqrt{2}} (|\Lambda S - \Sigma \rangle \mp |-\Lambda S \Sigma \rangle)  ] \notag \\
    =& \frac{1}{\sqrt{2}}\left[|+J_a\rangle |\zeta^\pm\rangle + |-J_a\rangle|\zeta^\mp\rangle\right]    \label{eq:wavefunction_Jx}
\end{align}
where 
\begin{equation}
    |\pm J_{a}\rangle = \frac{1}{\sqrt{2}}(|J\Omega M\rangle \pm |J-\Omega M\rangle)
\end{equation}
are the two nuclear rotational states whose $\hat{J}_a$ expectation values are 
\begin{eqnarray}
    \langle \hat{J}_a \rangle = \pm 1/2(J+1/2),
\end{eqnarray}
and
\begin{equation}
    |\zeta^\pm\rangle = \frac{1}{\sqrt{2}} (|\Lambda S - \Sigma \rangle \pm |-\Lambda S \Sigma \rangle)
\end{equation}
are the two electronic eigenstates that are symmetric with respect to the reflection through the plane $bc$, up to a phase factor.  According to Eq.~\ref{eq:wavefunction_Jx}, when the molecule has a positive angular momentum around the a-axis, the electronic ground state is dominantly $|\zeta^+\rangle$; when the molecule has a negative angular momentum around the a-axis, the electronic ground state is dominantly $|\zeta^-\rangle$.  This fact implies that, whereas the crossing phase space PESs in Fig.~\ref{fig:well} (copied in Fig.~\ref{fig:well_ef}\textbf{a}) are defined (and labeled) by their constant electronic character, the observed $e$/$f$ states arise and are defined (and labeled) by their relative energies as in Fig.~\ref{fig:well_ef}\textbf{b}.  In the latter case,  the $e$/$f$ states are not so much discrete states but rather arise as part of a continuous potential energy surface that must be requantized (as in Sec.~\ref{sec:2D_model} above). 

From a quantum chemistry perspective, if one considers the two BO electronic states of NO in one orientation as the degenerate "diabatic states" of interest, then the nuclear motion breaks this degeneracy, leading to the "adiabatic states" of interest in Fig.  \ref{fig:well_ef}.
However,  the $e$/$f$ states have identical energies at L = 0 because of the form of the splitting in Eq. \ref{eq:2D_en}, so that there are two different labeling schemes --  Fig.~\ref{fig:well_ef}a versus Fig.~\ref{fig:well_ef}b.  In a sense, one can argue that NO has a ``conical intersection''~\cite{yarkony1996diabolical,baer2006beyond,lehman2014dynamical} {\em in momentum space}, which provides a new interpretation of $\Lambda$-doubling (and a different view of the magnetic monopole derivative couplings discussed in Refs.~\citenum{moody1986realizations,bohm2013geometric}). Note that for the case of a diatomic molecule (as in Eq. \ref{eq:ci:sortof}), the conical intersection in phase space has co-dimension two (because only $L_a$ and $L_b$ enter the rotational Hamiltonian); more generally, in line with Ref.~\citenum{duston2025conical}, we would expect the conical intersection around $\bL=0$ to have co-dimension three.
Note also that our analysis above of Eq.~\ref{eq:2D_en} is effectively quantitative insofar as a purely quantum calculation using $L_a = L_b = \frac{1}{2}(J+\frac{1}{2})$ will necessarily lead to an energy splitting proportional to a factor   $ 2 \cdot \frac{1}{2}(J+\frac{1}{2}) \approx \sqrt{J(J+1)}$, which is in agreement with Eq.~\ref{eq:2D_en} above. 

\subparagraph{\REV{The $^2\Pi_{3/2}$ manifold}}
\REV{During the review process for this manuscript, a reviewer asked a very interesting question related to the scaling above.  Namely, the reviewer asked us if a phase space method would recover 
the correct form for  the $\Lambda$-splitting within the $^{2}\Pi_{3/2}$ manifold of NO.
The question above is interesting for two reasons. First, the origin of the $\Lambda$-splitting for the $^2\Pi_{1/2}$ manifold is different than the origin of $\Lambda$-splitting for the $^2\Pi_{3/2}$ manifold. For the former, the dominant contribution is the double perturbation of $H_\mathrm{SOC}^{(1)}$ and the Coriolis potential $H^{(1)}_\mathrm{Cor}$.  By contrast, for the latter, the lambda splitting
has  a different origin; here, the effect arises exclusively from the Coriolis potential, i.e. a second order perturbative treatment in $H^{(1)}_\mathrm{Cor}$ is required.  As a result of this difference, the lambda splitting scales as $J$ for the former (see Fig.~\ref{fig:2D_en}) and cubically for the latter. Obviously, one must wonder if a phase space approach can recover this difference? After all, if one were to follow Methods (i) and (ii) in Sec.~\ref{sec:3methods},  i.e. evaluating $\alpha$ at $\bP=0$  and then diagonalize a matrix as in  Sec.~\ref{sec:2D_model}, one would presumably find a splitting that is always linear  in $J$.
}

\REV{Second, a keen observer will note that, in order to recover the $\Lambda$-doubling results in Fig.~\ref{fig:2D_en} above, all that is required is that (i) one include SOC within a BO calculation and then (ii) one couple together the nuclear and electronic angular momentum through the $\hat{\bL}^e$ operator, as in Eq.~\ref{eq:E_use_Le} above.  In other words, one can in fact recover the $\Lambda$-doubling splitting of the $^{2}\Pi_{1/2}$ manifold without really ever invoking the phase space electronic structure Hamiltonian in Sec.~\ref{sec:review_PS}; if one wishes to avoid exploring phase space potential energy surfaces, one can model $\Lambda$-doubling directly with the relativistic BO states by including $\mathbf{\hat{L}_e}$ alone (and without building the $\hat{\mathbf{\Gamma}}$ operator).
That being said, just considering the scaling argument above, such an approach is clearly not appropriate for the $^2\Pi_{3/2}$ manifold. For the latter case, the only possibility is  
to invoke  Method (iii), fully explore the phase space potential energy surfaces, and then investigate the energy gap at $L=n \hbar/2$.
As  a result, such an approach will inevitably require that we build $\hat{\Gamma}$ for $\bP \ne 0$ (i.e. not just $\hat{\bL}^e$) and whether or not PS can recover the correct $\Lambda$-splitting for the $^2\Pi_{3/2}$ manifold is in fact a very strenuous test of the method.}

\begin{figure}[hbt!]
    \centering
    \includegraphics[width=0.5\linewidth]{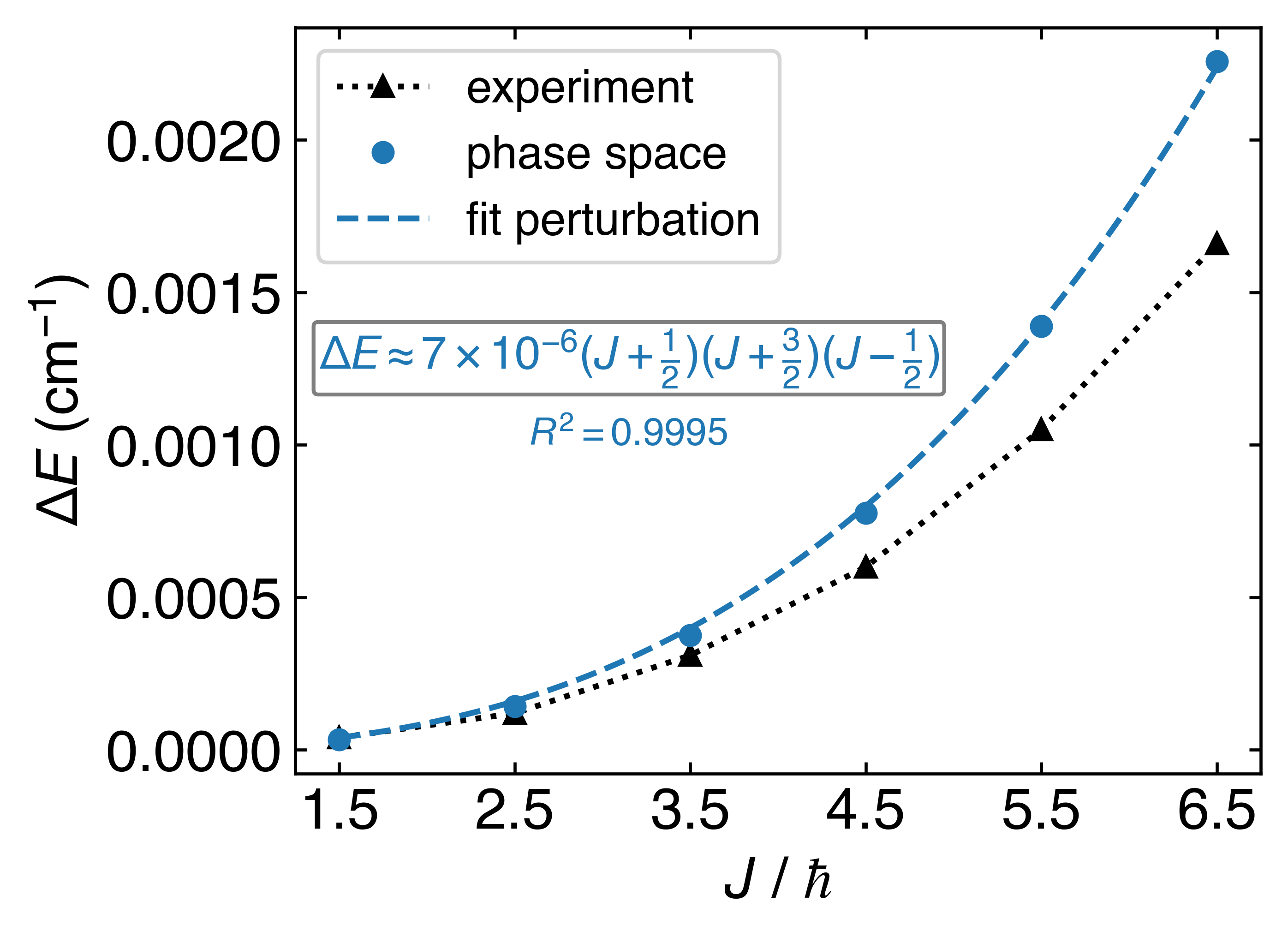} 
    \caption{\REV{Energy splitting between the $e$ and $f$ states of the $^{2}\Pi_{3/2}$ manifold as a function of total angular momentum $J$. For this data, unlike Fig.~\ref{fig:2D_en}, it is well known that the energy splitting requires a cubic (rather than linear) fit as a function of  $J$.  We compare phase space predictions, using a model parameter $\alpha$ derived from FCI with 1-electron SOC in the STO-3G basis with Method III (blue dots) versus experiment (black triangles). We further fit the ``phase space'' data to an analytic expression from standard perturbation theory, $\Delta E = k(J+\frac{1}{2})(J+\frac{3}{2})(J-\frac{1}{2})$, where $k$ is a fitted constant\cite{hinkley1972lambda} (dashed lines).  Note that, as in Fig.~\ref{fig:2D_en} above, our phase space predictions are quite accurate, which is another strong confirmation of a PS approach.}}
    \label{fig:3_half}
\end{figure}

\REV{As shown in Fig.~\ref{fig:3_half},  phase space theory does yield an accurate $\Lambda$-splitting as compared with experiment.  Note that, in order to resolve the tiny absolute energy scale (on the order of $10^{-10}$ Hartree), we required a tight convergence threshold for the diagonalization. In Fig.~\ref{fig:3_half}, we also fit our data to
an analytical energy expression from standard perturbation theory (valid under slow rotation and strong SOC)~\cite{hinkley1972lambda}, namely
$\Delta E \approx 7\times 10^{-6}(J+\frac{1}{2})(J+\frac{3}{2})(J-\frac{1}{2})$ cm$^{-1}$ (see Fig.~\ref{fig:3_half}, blue dashed curve).
Our results show a good fit, with a $R^2$ value of 0.9995.  Altogether, the data in Fig.~\ref{fig:3_half} again demonstrate the strong accuracy that can be found by solving electronic structure problems within a phase space framework.}

\paragraph{Conclusions and Outlook}
We have demonstrated that a simple phase space perspective on electronic structure theory provides a natural framework to capture the $\Lambda$-doubling in diatomic molecules. Thus, in line with previous discussions\cite{bian2025review,bradbury2025symmetry}, we assert that the same beyond-BO physics that underlies one of the smallest microscopic diatomic phenomena is also responsible for the macroscopic 
 Einstein de-Haas effect\cite{einstein1915experimenteller,mentink2019quantum, ganzhorn2016quantum} and, perhaps, chiral-induced spin selectivity\cite{bloom2024chiral, ray1999asymmetric, gohler2011spin}.  Interestingly, the major result here has been the nonadiabatic splitting  $\Delta E$ between different states of different angular momentum, where dynamically $\hbar/\Delta E$ corresponds to the tunneling time between degenerate states.  In the context of the Einstein de Haas effect, one is less interested in parity eigenstates (e.g. $e$/$f$), but rather in a basis with maximally aligned and opposite spin directions along the magnetic field direction. Nevertheless, the approach above should be able to isolate the tunneling time for a spin (or a collection of spins) to change orientation.  In other words, whereas  most treatments of Einstein de Haas phenomena use angular momentum conservation to predict the steady state nuclear rotational angular momentum that arises upon a total change in electronic 
spin angular momentum, the present approach based on phase space electronic structure theory will be able to go further and calculate transient effects: How long does it take for the spins to align? For the metal to reach a steady state of  rotation? And of course, in the long run, how much energy is lost if  we allow for frictional losses from nuclear vibrations?
 
With regards to this last point, although in this work we have restricted our attention to electron coupling to purely rotational nuclear motion, the most interesting physics undoubtedly lies ahead as we consider larger molecules and materials, and will need to model both vibrations and rotations -- and their coupled effect on electrons~\cite{burroughs2024infrared,davis2023infrared,davis2024bimolecular} -- all on an equal footing.  
That being said, in the future, if we seek to model larger molecules or materials, it will be essential to \REV{efficiently build the $\hat{\mathbf{\Gamma}}$  operators --  which will require some numerical tricks on our end and a more efficient implementation.  Our results here were calculated in PySCF\cite{sun2018pyscf,sun2020recent}, and our group also has a preliminary code in the Q-Chem package\cite{epifanovsky2021software}.} 
Our strong feeling is that, given the success of the current manuscript in tandem with previous results in VCD\cite{duston2024phase} and ROA\cite{tao2025non}, we will soon be equipped to explore a very broad set of questions: what are the roles of chiral phonons in determining spin-dependent electron transfer rates \cite{kim2023chiral}? How should we best quantitatively describe spin relaxation\cite{lunghi2019phonons}? Can we definitively prove the nature of the fundamental mechanisms behind different magnetic field effects\cite{steiner1989magnetic}?   
Lastly, this manuscript has suggested that whenever the ground state has a degeneracy built on different electronic angular momenta, nuclear motion will lead to a small splitting. As such, might not the equations above (especially Eq.~\ref{eq:ci:sortof}, which describes a gap opening due to off-diagonal electron-phonon coupling and is analogous to a Su–Schrieffer–Heeger (SSH)/Peierls Hamiltonian in the case of electron hopping\cite{su1979solitons, su1980soliton}) also apply to a metal and be a starting point for superconductivity calculations? When we move beyond a simple diatomic molecule, we can expect the number and the nature of interesting problems involving angular-momentum transfer to grow very rapidly.

\paragraph*{Acknowledgement}
The authors acknowledge helpful discussions with Xuecheng Tao, Robert W. Field, and Robert J. Gordon. This work was supported by National Science Foundation Grant No. CHE-2422858. NMK acknowledges support from the NSF CAREER award (CHE-2239624) and The Camille and Henry Dreyfus Foundation, Henry Dreyfus Teacher-Scholar Award (TH-24-019).

\bibliography{reference}

\end{document}